\documentclass[aps,showpacs,eps,preprint]{revtex4}
\usepackage{colordvi}
\usepackage{amssymb}
\usepackage{amsmath}
\usepackage{epsf}
\usepackage{mathrsfs}
\usepackage[dvips]{graphics}
\usepackage{graphicx, subfigure}
\usepackage{psfrag}
\usepackage[dvips,usenames]{color}
\begin{document}
\def \beq{\begin{equation}}
\def \eeq{\end{equation}}
\def \bea{\begin{eqnarray}}
\def \eea{\end{eqnarray}}
\def \bem{\begin{displaymath}}
\def \eem{\end{displaymath}}
\def \P{\Psi}
\def \Pd{|\Psi(\boldsymbol{r})|}
\def \Pds{|\Psi^{\ast}(\boldsymbol{r})|}
\def \Po{\overline{\Psi}}
\def \bs{\boldsymbol}
\def \bl{\bar{\boldsymbol{l}}}
\title{Electron transport and Goos-H\"anchen shift in graphene with electric and 
magnetic barriers: optical analogy and band structure} 
\author{Manish Sharma$^1$ and Sankalpa Ghosh$^{2,3}$}
\affiliation{$^1$Centre for Applied Research in Electronics, Indian Institute of Technology Delhi, New Delhi-110016, India}
\affiliation{$^2$Department of Physics, Indian Institute of Technology Delhi, New Delhi-110016, India}
\affiliation{$^3$Max-Planck Institut f\"ur Physik komplexer Systeme, N\"othnitzer Stra$\beta$e  38, 01187 Dresden, Germany}
\begin{abstract}
Transport of massless Dirac fermions in graphene monolayers is analyzed in the presence of  a combination of singular magnetic barriers and applied electrostatic potential.
Extending a recently proposed \cite{SGMS} analogy between the transmission of light 
through a medium with modulated refractive index and electron transmission in graphene through singular magnetic barriers to the present case, we find the addition of a scalar potential profoundly changes the transmission. We calculate the quantum version of the Goos-H\"anchen shift that the electron wave suffers upon being totally reflected by such barriers. The combined electric and magnetic barriers substantially modify the band structure near the Dirac point. This affects  transport near the Dirac point significantly and has important consequences for graphene-based electronics. 
\end{abstract}
\pacs{81.05.Tp,72.90.+y,73.23.-b,73.63.-b,78.20.Ci,42.25.Gy}{}
\date{\today}
\maketitle

\section{introduction}

In the ballistic regime, scattering of electrons by potential barriers can be understood in terms of phenomena like reflection, refraction and transmission, leading to an analogy between  electron transport and light propagation \cite{Book1, Datta}. Also, in a two dimensional electron gas (2DEG), now routinely produced in semiconductor heterostructures, it is well-established that transmission of de Broglie waves satisfying the Schr\"{o}dinger equation through a one-dimensional electrostatic potential is similar to light propagation through a refractive medium. This similarity can be used for lensing and focusing of electrons \cite{SIVAN90, Spector90}. Light propagation through optical fibres can also be understood in a similar manner \cite{GTS88}. 

For monolayer graphene, transport electrons do not follow the Schr\"{o}dinger equation  but instead behave as massless Dirac fermions leading to an intriguing set of transport phenomena \cite{KSN1, KSN2, YZ1, Geimreview, Castroneto, Beenakker}. To draw analogies with optics, graphene electrons must be described differently. Cheianov {\it et al}  \cite{Falko1} have shown that transport in graphene with an applied split gate voltage is akin to light propagating through a metamaterial with negative refraction index \cite{Veselago, Pendry}. Graphene electrons can also reflect from interfaces in a quantum version of the Goos-H\"{a}nchen effect  \cite{Beenakker2}. The possibility of guided modes in a graphene waveguide was  proposed recently \cite{Zhang09} and also studied for graphene constrictions in the sub-wavelength coherent transport regime \cite{Darancet}. 

Before the optical analogy can be developed, there is one other important consideration. While electrostatic potential barriers can manipulate transport in graphene, an electron  could tunnel through a high barrier in contrast to the conventional tunnelling of non-relativistic electrons \cite{Ando1, Falko, KSN3}. This behaviour, called Klein tunnelling, leads to several observable transport effects related to transport \cite{Levitov1}, some of which have been demonstrated in graphene \cite{Kim1,gordon} and also in carbon nanotubes\cite{Steele}. For practical graphene-based electronics, it is crucial to suppress Klein tunnelling so that electrons remain confined within a mesoscopic or nanoscopic size of the sample. It has been suggested that a magnetic barrier can do this \cite{MDE07} and many schemes have been proposed subsequently \cite{ZC08, Masir1, SGMS, Martino, Xu, Masir2, TKGhosh, Martino2, snake1, snake2, Park2}.

In this paper, we consider transport of massless Dirac fermions in graphene under the combined effect  of a magnetic barrier and an electrostatic voltage such as the one used in Ref.\cite{Falko}. The purpose is two-fold. One, graphene electrons have a linear band structure albeit only close to the Dirac point, coincident with the fermi level $E_F$ in undoped graphene. Small electrostatic potentials greatly affect electron states by shifting the Dirac point with respect to $E_F$ and causing the graphene sheet to behave as either an electron-deficit ($p$-type) or a hole-deficit ($n$-type) material. Thus, the effect of electrostatic potentials on any proposed graphene structures must be included. Two, we show how Klein tunnelling can be suppressed while transport can still be controlled using combined electric and magnetic barriers.  

The major findings of this paper are as follows.  Electron transport through combined electrostatic and magnetic vector potential (EMVP) barriers is explained using the language of geometrical optics. Transport through EMVP barriers is found to be substantially different than through MVP barriers studied earlier \cite{SGMS}. Tuning the electrostatic voltage effectively changes the magnetic barrier strength. Voltage can be tuned to a specific value at which zero modes of the modified Dirac operator are excited leading to highly asymmetric transmission. The  optical analogy is used to describe the Goos-H\"anchen shift, which can change sign as well as magnitude abruptly upon total internal reflection (TIR).  An analysis of transport through both finite and infinite series of EMVP barriers concludes our investigation.

\section{Electron transport through potential barriers and optical analogy}

We begin with a brief review of the optical analogues of non-relativistic and relativistic electron transport  through various electrostatic as well as magnetic barriers. We then discuss practical considerations for realizing such structures experimentally. 

\subsection{Theoretical framework}\label{recap}
When a non-relativistic electron in a 2DEG at fermi energy $E_F$ is incident on a potential barrier $V$, its momentum parallel to the interface outside and inside the barrier is conserved; i.e., $p_1 \sin \theta_1 = p_2 \sin \theta_2$, where $p_{1,2}$ are the momenta and  $\theta_{1,2}$ are the angles in the two regions. This leads to the following Snell's law \cite{SIVAN90}:
\beq \frac{\sin \theta_1}{\sin \theta_2} = (1-\frac{V}{E_F})^{\frac{1}{2}} \label{electronwave1} \eeq

For a 2DEG of Dirac fermions in undoped graphene, $E_F$ lies at the Dirac point where the conduction and valence bands touch. Thus, states near $E_F$ are equally populated by electrons and holes and the system is charge neutral. Application of $V$ locally lowers (raises) 
charge neutral Dirac point and $E_F$ lies in the conduction (valence)
band. Thus, $V$ can locally make a $n$ or $p$ type region 
in graphene and convert electrons into holes and vice versa inside a barrier. However, the conservation of chirality of fermions demands that $p_1 \sin \theta_1 = -p_2 \sin \theta_2$, giving a negative refractive index\cite{Falko, Falko1}.

To obtain a similar Snell's law for transport in the presence of a magnetic field, the field profile should scatter the electrons the same way as an electrostatic potential does. As magnetic fields bend electron trajectories continuously in a cyclotron motion, a direct analogy with light propagation is not possible as such. This consideration, however, changes in the presence of a highly inhomogeneous magnetic field. Particularly if the range of inhomogeneity is much smaller than the cyclotron radius, one is left with plane-wave like scattering states. For this to be valid, two conditions must be satisfied. One, the magnetic length $\ell_B=\sqrt{\frac{\hbar c}{eB}}$ should be similar in order to the width of such magnetic barriers. Second, the de Broglie wavelength $\lambda_F$ should be much larger than the width such that the electron will not see the variation in the vector potential inside the barrier.

An extreme case of such an inhomogeneous magnetic field is the one introduced in Refs.\cite{MPV94,magb2,periodic3,Nori1} having the following profile of the transverse magnetic field $\bs{B}$ and the corresponding vector potential $\bs{A}$ in the Landau gauge:
\bea \bs{B}& = & B_z(x)\hat{z}=B \ell_{B}[\delta(x+d)-\delta(x-d)]\hat{z} \nonumber \\
\bs{A}_y(x) & = & B \ell_B \Theta(d^2-x^2)\hat{y} \label{magbarrier} \eea
Such a magnetic field creates a wavevector dependent potential barrier which scatters electrons and  can be used for wavevector filtering. Recently \cite{SGMS}, we have shown that for  a  series of singular magnetic barriers \cite{MPV94}  fermions behave like light passing through an optical medium with a modulated refractive index \cite{yariv1}. However, the corresponding Snell's law is not specular as in classical optics. Changing direction or magnitude of the magnetic field of such barriers changes the refractive index.

It is then natural to ask whether electron scattering by such an MVP barrier admits an optical analogy similar to Eq.(\ref{electronwave1}). As discussed earlier, this would also confine electrons since Klein tunnelling will not occur through such inhomogeneous 
magnetic barriers.

Matching of momentum components and energy conservation for electrons scattered by magnetic barriers (cf. Eq.(\ref{magbarrier})) gives 
\beq \sin |\theta| = \sin |\phi| -\text{sgn}(\phi)\frac{1}{k_F \ell_B}, -\frac{\pi}{2} < \phi < \frac{\pi}{2} 
\label{angle1} \eeq 

Outside the barrier, $[k_x , k_y]$ are given by $k_F [cos \phi, sin \phi]$ and $\phi$ is the incident angle for an electron wave. Inside the barrier, $k_F[\cos \theta, \sin \theta]$ are given by $[q_x,k_y -\frac{1}{\ell_B}]$, where $\theta$ is the angle of refraction. The relation given in Eq.(\ref{angle1}) implies that, for a wave incident with positive $\phi$, the wavevector will bend towards the normal. Similarly, for a wave incident with negative incidence angle, the corresponding wavevector will bend away from the surface normal inside the barrier region.  A series of such magnetic barriers will thus lead to highly asymmetric transmission of electrons \cite{Masir1, SGMS}. 
Eq.(\ref{angle1}) also yields that when $|\sin |\theta|| > 1$ the angle $\theta$ becomes 
imaginary and TIR occurs. This naturally explains why such barriers can 
confine electrons. Confinement will occur when $\text{rhs(Eq.(\ref{angle1})}) > 1$ for  $\phi\in[\frac{\pi}{2}, 0)$ and  $< -1$ for  $\phi\in(0, \frac{\pi}{2}]$. In the latter case, this requires the wavevector to be negatively refracted 
at sufficiently high magnetic field before TIR occurs. 

For symmetric transmission, one could place two such single MVP barriers side by side but orient them oppositely \cite{SGMS} with the resultant magnetic field given by 
\bea \bs{B}&=&B_z(x)\hat{z}=B\ell_{B}[\delta(x+d)-2\delta(x)+\delta(x-d)]\hat{z}  \label{dmagbarrier}  \eea
Here, $B$ has an $x$ dependent vector potential pointing along the $y$-direction in Landau gauge. Energy conservation in medium $1$ ($-d < x < 0$) and medium $2$ ($0 < x < d$) leads to 
\bea q_{1,2}^2 + (k_y \mp \frac{1}{\ell_B})^2  =  k_F^2; \nonumber \\
\sin |\theta_{1,2}| =  \sin |\phi| \mp \text{sgn}(\phi)\frac{1}{k_F \ell_B } \eea
The incident angle is $-\frac{\pi}{2}<\phi<\frac{\pi}{2}$ and the angles of refraction are $\theta_1$ and $\theta_2$ in media $1$ and $2$ respectively. Thus, for such double MVP barriers, the wavevector bending towards (away from) the surface normal in the first half of the barrier bends away from (towards) the surface normal in the second half of the barrier achieving symmetric transmission. Because of TIR from the first as well as second half of a double MVP barrier beyond a critical angle of incidence, the reflectivity of such a barrier is relatively higher than that of a single MVP barrier. 
As was discussed in Ref.\cite{SGMS}, a series of such MVP barriers put side by side can work as a Bragg reflector, and the associated band structure also shows the effect  of the magnetic field on transport.

\subsection{Physical realization of EMVP Barriers}

\begin{figure}[ht!]
    \begin{center}
    \centerline{{\epsfxsize 8cm  \epsffile{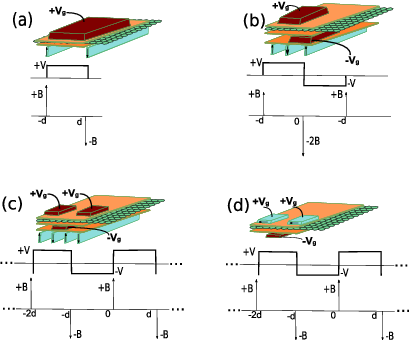}}}
    \end{center}
    \caption{%
        {\it (color online)}  EMVP barrier structures for graphene. In (a)-(c), the magnetic field is applied using patterned ferromagnetic (FM) lines with perpendicular anisotropy and two potentials $+V_g$ and $-V_g$ are applied by separate conductor lines. As given in the text, effective potential induced in the graphene is $V$. In (b), the magnetic strength of the middle line is doubled by either using a different FM material or by larger dimensions. In (d), the magnetic field is produced from the two edges of a FM stripe with in-plane anisotropy and the same stripe is also used to apply  one of the potentials.
     }%
         \label{ferromag}
\end{figure}

 In Fig.\ref{ferromag} are depicted possible structures that could be made. A graphene sheet is placed in close proximity to long magnetic stripes that produce delta function-like magnetic fields. Each of the planes of conductors, FM stripes and graphene are separated by insulating dielectric layers. Typically, field patterns are confined to within a few ten's of nanometers and it is possible to make stripes at various length scales where the above-mentioned delta function approximation works well.  
 
It is important to discuss if the nanostructures required for producing the desired EMVP profiles are experimentally realizable. The two requirements are the appropriate gate voltages and the B fields. The dielectric films needed to sandwich graphene and apply the gate voltage $V_g$ can be made as thin as several nm's. The actual voltage seen by the graphene layer is an effective voltage $V$ as discussed later. Since the typical breakdown strength of dielectrics such as alumina and fused silica are around $10-20 MV/m$, a graphene layer could be subjected to 1V applied across a 100nm thick dielectric. Following Refs.\cite{KSN1,KSN2}, gate voltages $V_g$ of upto $\pm100 V$ have already been applied to graphene flakes.  The electrostatic gate potentials would be generally applied by separate conductors placed suitably in different planes than the FM stripes, but in some cases, the FM stripes could also be used to apply voltages.

The B field profiles required can be generated using demagnetizing fields produced at the edges of narrow stripes made with hard ferromagnetic (FM) materials of either perpendicular (Figs.\ref{ferromag}(a)-(c)) or in-plane anisotropy (Fig.\ref{ferromag}(d)). The major component of the demagnetizing field reaches the graphene monolayer and produces the desired field profile. Though there is always some component of the demagnetizing field that will give rise to undesired fringe fields in other directions, these can be substantially lowered by suitable magnetic design of the stripes. Such nanostructures are routinely used in magnetic recording media \cite{appphys}. Materials such as CoCrPt  produce fields of 1 Tesla close to the surface with bit lengths ranging from 50-100nm. Stripes down to 10 nm can also be patterned by nanolithography \cite{terriskrawczykbader}. Isolated double magnetic barriers or in a periodic pattern can also be created in one (1-D) or both (2-D) dimensions, although in this paper we shall only discuss 1-D structures. Magnetic field patterns over various length scales can be obtained. As an example \cite{IEEE}, isolated tracks of single-domain magnetic islands have been fabricated using focused ion-beam lithography for both perpendicular and parallel anisotropy using CoCrPt.  Here, patterns with successive magnetizations pointed along opposite directions were achieved. In another recent work, off-axis electron holography has been used to probe the magnetization structure in high density recording medium by using perpendicular magnetic anisotropic (PMA) recording medium \cite{dipole}. The direct imaging of magnetization done shows that the foils of PMA material consist of successively reversed highly stable domain structures of few ten's of nanometer size. In practice, one can also change the strength of the magnetic field by suitable adjusting the width of such PMA material. Precise design of read-write structures for recording individual bits at these dimensions has also been achieved \cite{nphotonics}. Using the above discussed techniques, typical magnetic barriers can be patterned down to 50-100nm widths ($d$ in Fig.\ref{ferromag}).

In a recent experiment by S Pisana et al.\cite{Pisana}, the enhanced
magneto-resistance of a monolayer graphene sheet has been measured by
connecting it to two voltage and two current terminals and simultaneously
exposing it to various magnetic field strength at room temperature. The
differential voltage as a function of the magnetic field has been
plotted. From the data the joint effect of  magnetic field and the applied
voltage on the magnetotransport particularly close to the Dirac point has been
analyzed. From the analysis of the magneto-resistance data 
it was inferred that the band structure of transport gets
strongly modified in presence of voltage + magnetic field composition. A
direct comparison with the experimental data is difficult since the
magnetic field profile in that experiment is homogeneous and the voltage
and current probes used also lead to a different geometry. However the
above experiment result clearly shows that simultaneous application of
voltage and magnetic field strongly influences band structure 
pointed out in this work. Recently Girit et al. \cite{Girit} fabricated a SQUID in graphene,
another practical device structure patterened with lithography, and showed that electronic transport
can be tuned by applying a magnetic field and a voltage bias. 

All these examples clearly show that the structures we have proposed alongwith the appropriate electric and magnetic fields are very much in the realm of current technology and experimentally realizable. 

\section{Transport in EMVP Barriers}\label{secIII}

The electron focusing property of a single layer of graphene due to an electrostatic barrier has already been identified \cite{Falko1}. In this section, we shall see how this behaviour gets modified in the presence of a combined electrostatic and MVP (EMVP) barrier. We shall also discuss transport through several simple but illustrative types of EMVP barriers.

\subsection{Single EMVP barrier}\label{emvp1}

For undoped graphene, $E_F$  lies at the charge-neutral Dirac point and the quasiparticles behave like massless Dirac fermions obeying linear dispersion. A gate voltage $\pm V_{g}$ can be applied using a metal electrode and separating the electrode from the graphene layer  with an insulating oxide layer as discussed earlier in Fig.\ref{ferromag}(a). When such a voltage $V_g$ is applied  locally it proportionally induces electron (hole) doping $\pm \sigma_n$. This induced charge locally shifts the undoped Fermi level $E_{F}$ from the Dirac point by an amount $V=\text{sgn}(\sigma_n)\sqrt{\sigma_n} \hbar v_{F}$ where $\text{sgn}{(\sigma_n)}$ is the sign of the induced charge. This creates a local  potential barrier $V$, which gives the difference between the local fermi level and the fermi level in the undoped region.  Henceforth, by voltage $V$ we shall mean the height of this local potential barrier and not the actual gate voltage $V_g$.

It is relevant to discuss by how much can the Dirac point actually be shifted while the dispersion still remains linear for not too high $V$.  Typically, graphene devices are  characterized by electric field measurements \cite{KSN1, KSN2, YZ1} which gives a zero bias doping ($V_{g}=0)$ of order $10^{12} cm^{-2}$. 
Following \cite{KSN1,KSN2}, it can be argued that this effective massless Dirac fermion characterization is well valid as the gate voltage $V_{g}$ is varied between $\pm 100 V$. Also, in recent experiments the energy dispersion of Dirac fermions at Brillouin zone corners was directly measured for graphite \cite{MDF}. The energy range over which the measurement was carried out, namely $E-E_{F}$ is of the order of a few volts. Even though a direct comparison of our result is not 
accurate with such experiments, the typical values for the potential barrier $V$ in this paper have been decided by keeping in mind this limit.

\begin{figure}[ht]
\centerline{{ \epsfxsize 7cm \epsfysize 5.5cm \epsffile{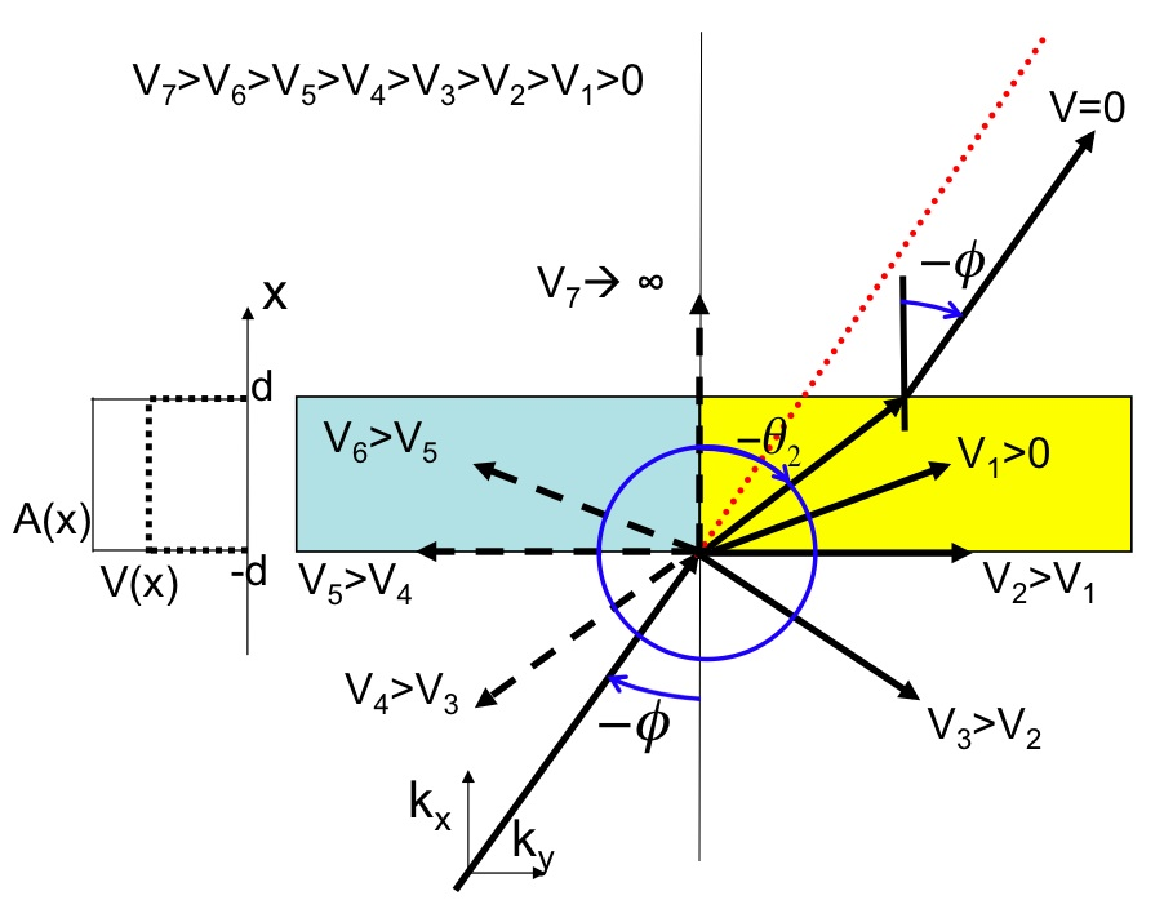}}}
\caption{{\it (color online)} Propagation through a single EMVP barrier. 
With increasing $V$, the refracted angle $\theta_2$ increases continuously 
in the clockwise direction. Dashed rays are negatively refracted.}
\label{figraydiagram}
\end{figure}

We consider the transmission at incident energy $E$ set to $E_F$ through a single EMVP barrier of the form
\bea V(x) =  V~\text{with}~ V >0;  A_{y}(x) &=& B \ell_B,~  |x| < d \nonumber \\
     V(x) = 0; A_{y}(x) &=& 0,~~~~~ |x| > d .  
\eea  
It is assumed that $B$ is not strong enough to break the degeneracy of the $K$ and $K'$ points. As discussed earlier, for both $\pm \text{ve}$ voltage the Dirac point is shifted with respect to $E_F$ and this creates local n- or p- doped regions. For a purely magnetic barrier, there are no separate n- or p- regions \cite{SGMS}. At either $K$ or $K'$, the motion is described by
\beq  v_F \begin{bmatrix} \frac{V}{v_F}  & \hat{\pi_x} - i \hat{\pi_y} \\
     \hat{\pi_x} +  i \hat{\pi_y}   & \frac{V}{v_F} \end{bmatrix}
     \begin{bmatrix} \psi_1 \\ \psi_2 \end{bmatrix} = E \begin{bmatrix} \psi_1 \\ \psi_2 \end{bmatrix}\label{diracemvp},~-d \le x \le d \eeq
Here, $\hat{\pi}=\hat{p} + \frac{q}{c}\vec{A}$ and $V$ implies $qV$ where 
$q=-|e|$. Since the vector potential is in Landau gauge, the stationary solutions can be written as    
\beq \psi_{1,2}(x,y) = \phi_{1,2}(x) e^{ik_y y} \nonumber \eeq 
Substituting these solutions in Eq.(\ref{diracemvp}), one gets the coupled one-dimensional equations 
\beq \begin{bmatrix}  0 & -i\partial_x -i(k_y - \frac{1}{\ell_B}) \\
      -i\partial_x +i(k_y - \frac{1}{\ell_B})  & 0  \end{bmatrix}
     \begin{bmatrix} \phi_1 \\ \phi_2 \end{bmatrix} = \frac{E-V}{\hbar v_F} \begin{bmatrix} \phi_1 \\ \phi_2 \end{bmatrix}, \label{diracemvp2}\eeq
which can be decoupled to yield 
\beq [ -\partial_x^2 + (k_y  -\frac{1}{\ell_B})^2 ] \phi_{1,2} = [\frac{E-V}{\hbar v_F}]^2 \phi_{1,2} 
\eeq 
The corresponding stationary solutions $\phi_{1,2}(x)$ 
are 
\beq
\phi_1 = \left\{
\begin{array}{rl}

e^{ik_x x} + r e^{-ik_x x} & \text{if }~~ x < -d\\
a e^{iq_x x} + be^{-i q_x x}  & \text{if }~~ -d < x < d \\
t e^{i k_x x} & \text{if}~~ x > d
\end{array} \right.
\label{particle1}\eeq
\beq \phi_2 = \left\{
\begin{array}{rl}
s[e^{i (k_x x +  \phi)} - r e^{-i(k_x x + \phi)}] & \text{if }~~ x < -d\\
s'[a e^{i(q_x x + \theta)} - b e^{-i(q_x x + \theta)}] &   \text{if }~~ |x|<d 
\\ 
s te^{i(k_x x +\phi)} & \text{if }~~ x > d
\end{array} \right . 
\label{hole1}
\eeq
These are similar in form to those for a pure magnetic barrier \cite{SGMS} or an electrostatic step potential \cite{KSN3}. But 
$\{ k_x,k_y \}$ and $\{q_x, k_y - \frac{1}{\ell_B} \}$, namely the  $x$ and $y$ components of the wavevector,  inside and outside the barrier regime are  different. Here, $s,s'$ are $sgn(E-V)$ in the respective regions. Upon setting incident energy $E_F$ as $\hbar v_F k_F$, 
substitution of  Eqns.(\ref{particle1}) and (\ref{hole1}) in Eq.(\ref{diracemvp2}) leads to 
\bea k_x^2 + k_y^2 & = & k_F^2,~\text{with}~k_F=\frac{E_F}{\hbar v_F}
~ |x| > d \nonumber \\
q_x^2 + (k_y -\frac{1}{\ell_B})^2 & = & (k_F - \frac{V}{\hbar v_F})^2 
=k_F'^2, ~~ |x| \le d 
\label{energyemvp}\eea
The incidence angle $\phi$ and the refraction angle $\theta$ are given by  $\tan^{-1}(\frac{k_y}{k_x})$  and $\tan^{-1}(\frac{k_y\ell_B - 1}{q_x\ell_B})$ respectively. 
Eq.(\ref{energyemvp}) can then be rewritten to obtain the Snell's law analogue for electron waves of such massless Dirac fermions incident on the EMVP barrier as
\bea \sin |\theta| & = & 
S_F( \sin |\phi| - \text{sgn}(\phi)\frac{1}{k_F \ell_B}) \label{angle11}  \\
& = & S_F \sin |\theta||_{V=0} \nonumber \\
S_F & = & \frac{k_F}{k_F'}=s|\frac{E_F}{E_F-V}| ; ~s=sgn(E_F-V)  \nonumber \\
\nonumber \eea
 Comparison between Eq.(\ref{angle1}) and Eq.(\ref{angle11})  shows that 
the potential barrier effectively scales the 
refraction angle by the scale factor $S_F$ defined above. $S_F$ is a non-monotonic and discontinuous function of $V$ and we shall study its 
impact on the refraction of the incident electron wave. 

For positive incidence angles $\phi \in (0,\frac{\pi}{2} )$ and $V$ below $E_F$, $S_F$ is positive and increases with $V$. For not too high $B$ and
not too low $\phi$, $\sin \theta|_{V=0}$ is positive and less than $\sin \phi$. 
Thus, upon multiplication by $S_F$, $\sin \theta$ and $\theta$ increases with 
$V$.Thus, the refraction angle is larger than the refraction angle for pure magnetic barrier ($V=0$). As a result, the wavevector bends increasingly away from the surface normal.

For $V=0$, electrons are going from a rarer to a denser medium. With increasing $V$, $\theta$ will increase for a constant $\phi$ and electrons behave like passing into an increasingly rarer medium. At some point, $\sin\theta$ becomes greater than $\sin \phi$, making the barrier regime behave like a rarer medium as compared to the region outside. 

If $V$ is such that rhs of Eq.(\ref{angle11}) is greater than $1$, the electron wave suffers TIR at this junction. Thus, the following result implies
that for a given strength $B$ of MVP barrier and a given angle of incidence $\phi$, by increasing $V$, it is possible to totally reflect the electron wave. Since the reflectivity of a magnetic barrier increases with higher $B$, this implies that the addition of  $V$ can effectively convert a weaker magnetic barrier into a stronger one. 

The point $V=E_{F}$ represents a singularity in the spectrum and demands a separate discussion. In the absence of magnetic barriers, such a point 
represents the zero modes for Dirac operators and leads to the emergence of new Dirac points. This has been discussed in a number of recent works considering Andreev Reflection in a graphene based NIS junction 
\cite{Shubhro} and for electronic states of graphene in a periodic potential \cite{Brey, CHPark2, CHPark3}. The presence of a magnetic barrier breaks the time reversal symmetry explicitly and these zero modes become the zero modes of the modified Dirac operator and the corresponding solutions are different. The equations satisfied by $\phi_{1,2}$ are 
\beq [-\partial_x^2 + (k_y - \frac{1}{\ell_B})^2]\phi_{1,2} = 0 \label{zeromodes} \eeq
Thus, the solutions along the $x$-direction are either exponentially 
decaying or growing. Since these solutions exist in the region of TIR we shall retain only the decaying one. 
This can be contrasted with the case without any magnetic barrier \cite{Brey} where the equation obeyed by the zero  mode solution is $-\partial_x^2 \phi_{1,2}=0$, yielding linear solutions instead of an exponential one. The significance of such zero modes is that 
on the two sides of the singular point,
the relative sign between the $\phi$ and $\theta$ becomes opposite. 

As $V$ is increased further beyond $E_F$, $S_F$ becomes negative 
and increases from $-\infty$ to $0$. This implies
that the relative sign between the angle of incidence and refraction will 
remain opposite.
For this  range of $V$, the $EMVP$ barrier acts like a left-handed metamaterial with negative refraction properties. As long as rhs of Eq.(\ref{angle11}) remains less than $-1$   
the electron wave is totally internally reflected. However, at very high $V$,  $|S_F|$  becomes less than $1$. The electron wavevector will again retrace its path back to the first medium while remaining negatively refracted. As $V$ approaches $\infty$ the refraction angle becomes almost zero allowing the electron beam propagating along the surface normal through negative refraction. 

For negative incidence angles, namely  $\phi \in (-\frac{\pi}{2},0 )$, upon setting $V$ to $0$, one gets an MVP barrier \cite{SGMS}. Here,  the refraction angle  $|\theta|$ is larger than incidence angle $|\phi|$. Thus, electrons are seen as passing from a denser to a rarer medium. When $V$ is turned on, $|\theta|$ increases continuously till $V$ reaches $E_F$ and eventually the electron wave suffers TIR. Thus, again by increasing $V$ it is possible to totally reflect an electron wave for any given $\phi$ and $B$. When $V$  surpasses $E_F$, the sign of $\sin |\theta|$ will be opposite to the sign of $\sin |\phi|$. At smaller $V$ close to $E_F$ there is still TIR, whereas for very high $V$ much above $E_F$ the refraction becomes negative and the wave again retraces its path back in the barrier regime. 
This situation is depicted in Fig.\ref{figraydiagram}, where  the wavevector  for the refracted ray is shown changing with increasing $V$.

\begin{figure}[ht]
\centerline{{ \epsfxsize 7.5cm \epsffile{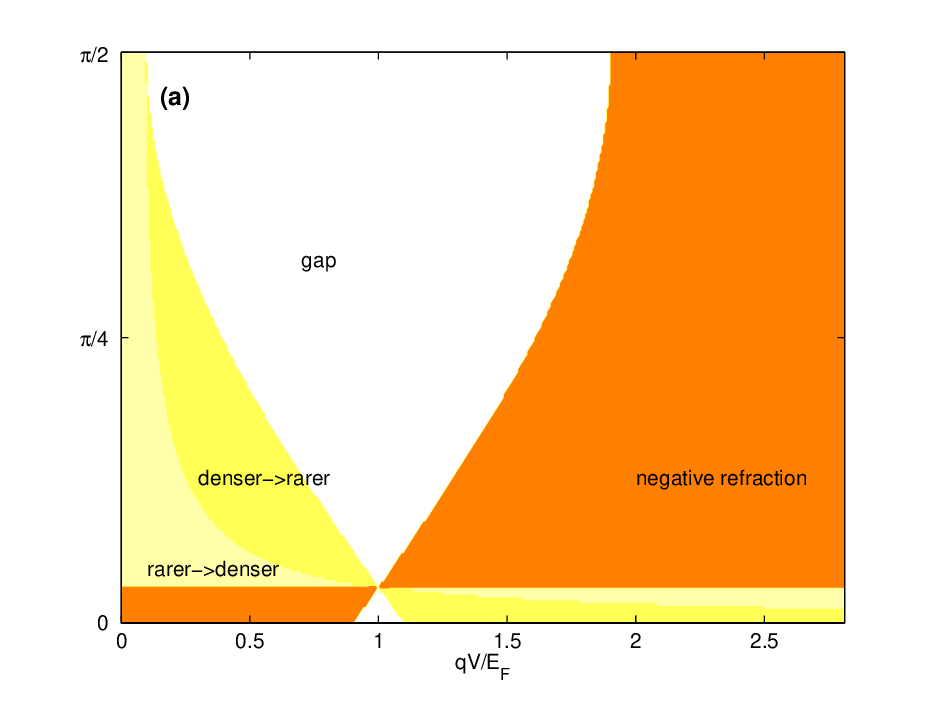} }}
\centerline{{\epsfxsize 7.5cm \epsffile{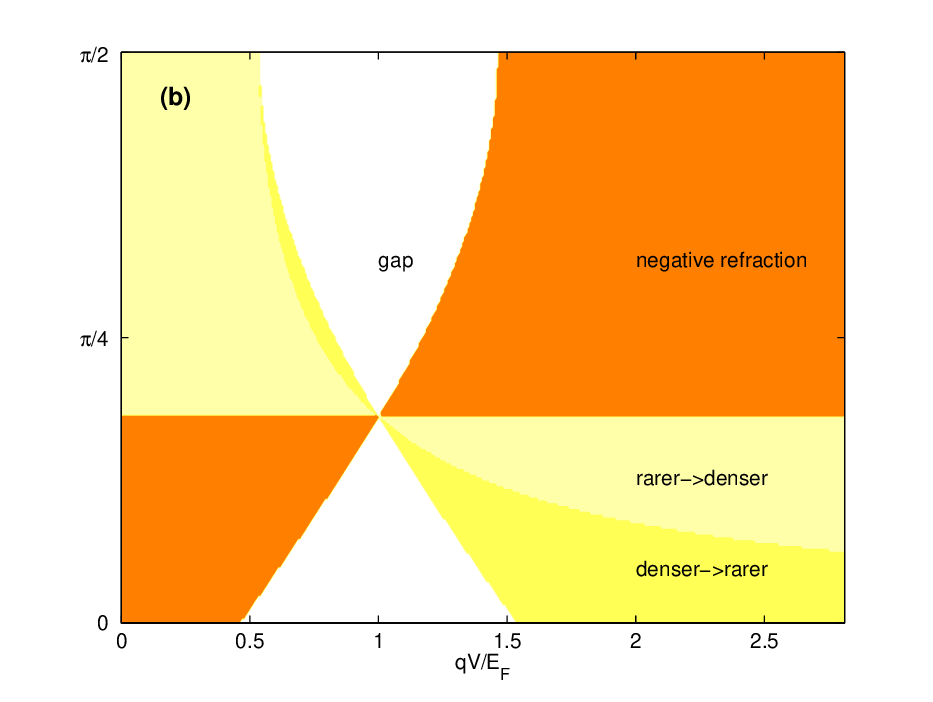}}}
\caption{{{\it (color online)} Phase diagram of
refraction angle $\theta$ at magnetic field $0.1$ Tesla and $3$ Tesla. The gap corresponds to the region
where TIR occurs. $d=100nm$ for this as well as for all subsequent figures. }}
\label{figtheta11tesla}
\end{figure}

The preceding discussion is  summarized by plotting $\theta$ as a function of the incidence angle $\phi$ for different $V$ and $B$ values in Fig.\ref{figtheta11tesla}.
It is important to note each of Fig.\ref{figtheta11tesla}(a) and (b) are separated into an upper and a lower part separated by the value $\phi_s = \sin^{-1}(\frac{1}{k_F \ell_B})$. In the upper part where $\phi$ is larger than $\phi_s$, with increasing $V$ the refracted angle $\theta$ exhibits four phase regions: rarer $\to$ denser, denser $\to$ rarer, TIR gap, negative refraction. In the lower part, the four phase regions exhibit a different order with increasing $V$:  negative refraction, TIR gap, denser $\to$ rarer, rarer  $\to$ denser. All these different regions meet at the limiting point $(E_F, \phi_s)$, where the behavior is singular and is described by Eq.(\ref{zeromodes}). 

%

\begin{center}
\begin{figure*}[ht]
\centerline{{ \epsfxsize 15 cm \epsffile{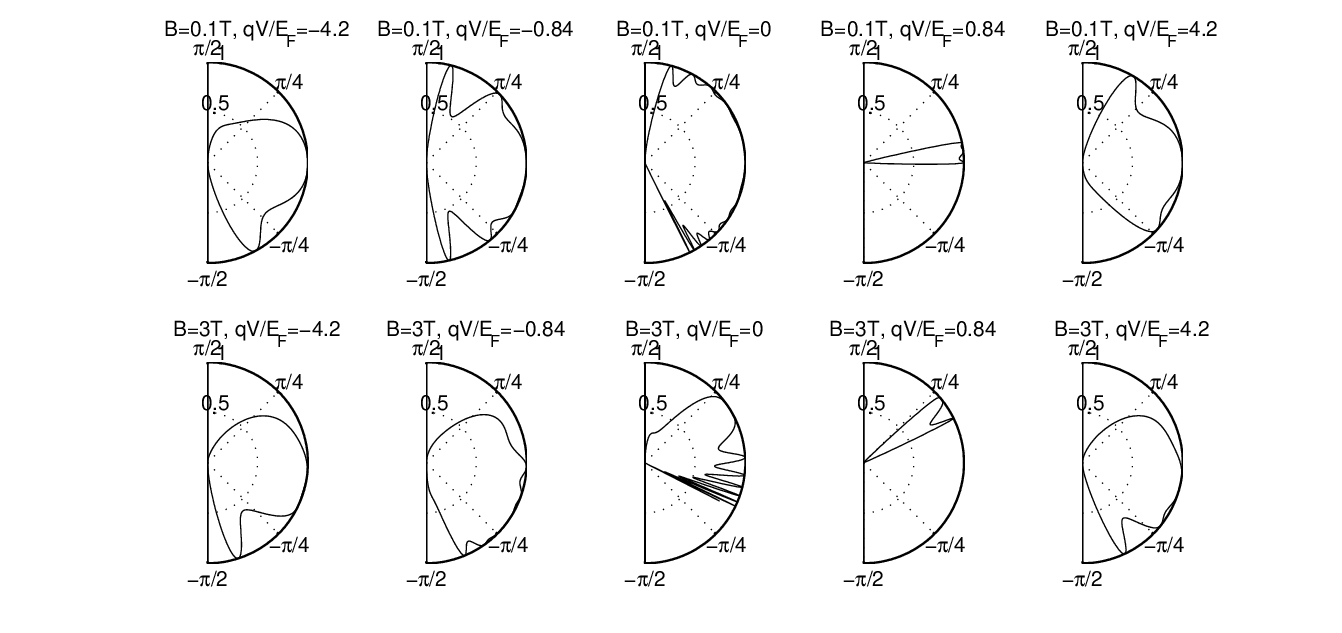}}}
\caption{{Transmittance $T$ through a single EMVP barrier at $B=0.1$ Tesla and $3$ Tesla at different angles of incidence $\phi$.}}
\label{transEMVP}
\end{figure*}
\end{center}

Transmission through a magnetic barrier gets strongly affected by $V$. By continuity at $x=\pm d$  in Eqs.(\ref{particle1}) and (\ref{hole1}), the actual dependence can be obtained as 
\beq t = \frac{2ss'e^{-ik_{x}D}\cos\phi\cos\theta}{ss'[e^{-iq_xD}\cos(\phi+\theta)+e^{iq_x D}\cos(\phi-\theta)]-2i\sin{q_xD}}, \label{temvp} \eeq
where $D=2d$ and now $q_x=k_{f}^{'}\cos\theta$.  
The transmittance  ($T=t^{*}\cdot{t}$) and reflectance  $R=1-T$ are of the same form as given in Ref.\cite{KSN3}. In Fig.\ref{transEMVP} is  plotted $T$  as a function of $\phi$ and $V$.  The central column depicts the case for with $V=0$, same as  MVP barriers \cite{SGMS},  for comparison. 

At the singular point $qV=E_F$, the barrier then becomes fully reflecting.
To study the transmission properties as this singular point is approached,
$qV$ is increased to $0.84E_F$ (third column plots). In this case, transmission takes place over a very small window along the $\phi$ axis asymmetrically located in one quadrant. When $B$ is changed from $0.1T$ to $3T$, a similar narrow window of transmission gets even more shifted to one quadrant of the $\phi$ axis. This is because the deviation from the normal direction $\frac{1}{k_F \ell_B}$ increases with $B$. 

On the other hand, if $qV$ is changed to the opposite sign, namely $-0.84E_F$, $S_F$ is always a fraction. Thus, $\theta$ on both sides of the surface normal will now be decreased upon increasing $V$ and the critical angle for TIR will go up. Consequently, we see that for the same magnitude but of opposite sign of $V$ transmission exists over a large range of $\phi$ in the second column plots.

If $|qV|$ is increased to $4.2E_{F}$ such that $|qV| \gg E_F$, the barrier becomes transmitting for all $\phi$. This can be seen in the first and fifth column plots. This also agrees with the refraction map in Fig.\ref{figtheta11tesla}(a), where at higher $V$ along most of the regions of the $\phi$  axis, $\theta$ is real. In the lower row of  $B=3T$, the higher magnetic field also means a much higher reflectivity, and the effect of turning on an electric field is to introduce more asymmetry in the transmission as one compares $T(\phi)$ with $T(-\phi)$. 

To summarize, the above analysis shows the transmittance  through EMVP barriers can be controlled from very low to very high  value  
by switching $V$ from positive to negative for a fixed strength $B$ of pure MVP barriers \cite{SGMS} leading to a very high degree of rectification. This constitutes one of the major results in this paper.

\subsection{A double EMVP barrier}\label{doubleEMVP}

We now consider a double MVP barrier added with a split gate voltage, henceforth called a DEMVP barrier. The vector and scalar potentials are characterized by 
\bea  V(x) & = & -V; A_y(x) =B\ell_B, -d  < x < 0:~\text{region I} \nonumber \\
      V(x) & = & V ;   A_y(x) =-B \ell_B, 0 < x < d:~\text{region II}~\label{split-gate} 
\eea

The schematic diagram in Fig.\ref{ferromag}(b) shows how such DEMVP barriers can be constructed by placing metal electrodes and insulating layers between PMA materials with alternating magnetizations and the graphene sheet. 

In regions I and II, the charge neutral 
Dirac point is raised  and lowered with respect to 
$E_F$. Thus, regions $I$ and $II$ become $p$ and $n$ type
and create a p-n junction on monolayer graphene. Similarly, one can  switch a p-n junction to an n-p junction by reversing voltage, an effect which can be used to make structures for electron focusing \cite{Falko}. We now look at the effect of highly localized magnetic barriers on the transport through such a p-n (n-p) junction. Again with the incident electron energy $E$ set to $E_F$ and assuming $|V|$ is not too high to break the Dirac fermion approximation, we get 
\beq  v_F \begin{bmatrix} \pm \frac{V}{v_F}  & \hat{\pi_x} - i \hat{\pi_y} \\
     \hat{\pi_x} +  i \hat{\pi_y}   & \pm \frac{V}{v_F} \end{bmatrix}
     \begin{bmatrix} \psi_1 \\ \psi_2 \end{bmatrix} = E_F \begin{bmatrix} \psi_1 \\ \psi_2 \end{bmatrix},
      \label{pnmag} \eeq
where the $-$ve sign is for region I  and the $+$ve sign is for region II.
The stationary solutions in Landau gauge are $\psi_{1,2}(x)=\phi_{1,2}(x)e^{ik_{y}}$, where the subscripts $1,2$ denote regions I,II respectively.
Substitution of these solutions in Eq.(\ref{pnmag}) gives 
\bea \hbar v_{F} [-i \frac{\partial}{\partial x} 
- i (k_y \mp \frac{1}{\ell_B})]\phi_2 
& = & (E_F \pm V) \phi_1 \nonumber \\
\hbar v_{F}  [-i \frac{\partial}{\partial x} + i (k_y \pm \frac{1}{\ell_B})]\phi_1 & = & (E_F\pm V)\phi_2 \label {DEMVPeq} \eea
Here, the first and second signs are for regions I and II respectively.
Each pair of the above equations can now be decoupled to yield
\bea (\hbar v_{F})^2 [-\frac{\partial^2}{\partial x^2} + 
(k_y -\frac{1}{\ell_B})^2]\phi_{1,2} & = & (E_F+V)^2\phi_{1,2}, \nonumber \\ 
(\hbar v_{F})^2 [-\frac{\partial^2}{\partial x^2} + 
(k_y + \frac{1}{\ell_B})^2]\phi_{1,2} & = & (E_F-V)^2\phi_{1,2},   
\label {ekDEMVP} \eea 
The stationary solutions that satisfy the above equations are left and right moving plane waves of the form 
\beq
\phi_{1,2} \propto \left\{
\begin{array}{rl}
e^{iq_1 x},&\text{if }-d< x < 0 \\
e^{iq_2 x},&\text{if}~0< x < d 
\end{array} \right.
\label{basis}\eeq
Since $E_F$ is $\hbar v_F k_F$,
\bea k_x^2 + k_y^2  = & k_F^2, &\text{if}~|x| > d \label{dem1} \\
     q_1^2 + [k_y - \frac{1}{\ell_B}]^2 = & [\frac{E_F+V}{\hbar v_F}]^2=
[k_F^{1}]^2, &\text{if}~-d < x \le 0 
\label{dem2} \\
q_2^2 + [k_y + \frac{1}{\ell_B}]^2 = & [\frac{E_F-V}{\hbar v_F}]^2=
[k_F^{2}]^2, &\text{if}~ 0 < x \le d  
\label{dem3}\eea 
Here, $k_F^{1,2}$ are the modified wavevectors in regions I and II and  the explicit solutions in the various regions will be
\beq
\phi_1 = \left\{
\begin{array}{rl}
e^{ik_x x} + r e^{-ik_x x} & \text{if } x < -d\\
ae^{iq_1 x} + be^{-i q_1 x}  & \text{if } -d < x < 0 \\
ce^{iq_2x} + de^{-iq_2x} & \text{if} 0 < x < d \\
t e^{i k_x x} & \text{if} x > d
\end{array} \right.
\label{particle2}\eeq
\beq \phi_2 = \left\{
\begin{array}{rl}
s[e^{i (k_x x +  \phi)} - r e^{-i(k_x x + \phi)}] & \text{if } x < -d\\
s_1[a e^{i(q_1 x + \theta_1)} - b e^{-i(q_1 x + \theta_1)}] &   \text{if } d< x <0 \\
s_2[c e^{i(q_2 x + \theta_2)} - d e^{-i(q_2 x + \theta_2)}] & \text{if}  0 < x < d \\
s te^{i(k_x x +\phi)} & \text{if } x > d
\end{array} \right.
\label{hole2}\eeq
with \beq \tan \theta_{1,2}=\frac{k_y \mp \frac{1}{\ell_B}}{q_{1x,2x}}, s_{1,2}=\text{sgn}(E_F \pm V) \nonumber \eeq 

The angle  
$\theta_{1,2}$ gives the angle between the propagation vector and $x$-axis in regions I,II. Here, $s_{1,2}$ can be $\pm 1$ depending on $V$ for a given $E_F$. Using the definition of $\theta_{1,2}$ and Eqs.(\ref{dem1}) -(\ref{dem3}),
\bea \sin \theta_1 & = & 
S_F^{1}[\sin \phi - \text{sgn}(\phi)\frac{1}{k_F \ell_B}] \label{theta1} \\
\sin \theta_2 & = & S_F^{2}[\sin \phi + \text{sgn}(\phi)\frac{1}{k_F \ell_B}] \
\label{theta2} \\
S_F^{1,2} & = & \frac{E_F}{E_F \pm V}= \frac{k_F}{k_F^{1,2}} \nonumber \eea

Thus, $\sin \theta_{1,2}$ for the double MVP barrier
$(V=0)$ gets multiplied by the scale factor $S_F^{1,2}$ to yield $\sin \theta_{1,2}$ for a DEMVP barrier for a nonzero $V$ creating the split gate voltage. Because of the sign reversal, $S_F^1$ and $S_F^2$ behave in different ways. While $S_F^{2}$ behaves 
in the same non-monotonic singular way as $S_F$ 
for a single EMVP barrier, $S_F^{1}$ decreases monotonically from $1$ to $0$ as $|V|$  increases from $0$ to $\infty$. For $V <0$, the effect just gets reversed. 

\begin{figure}[ht]
\centerline{{ \epsfxsize 7.5cm \epsffile{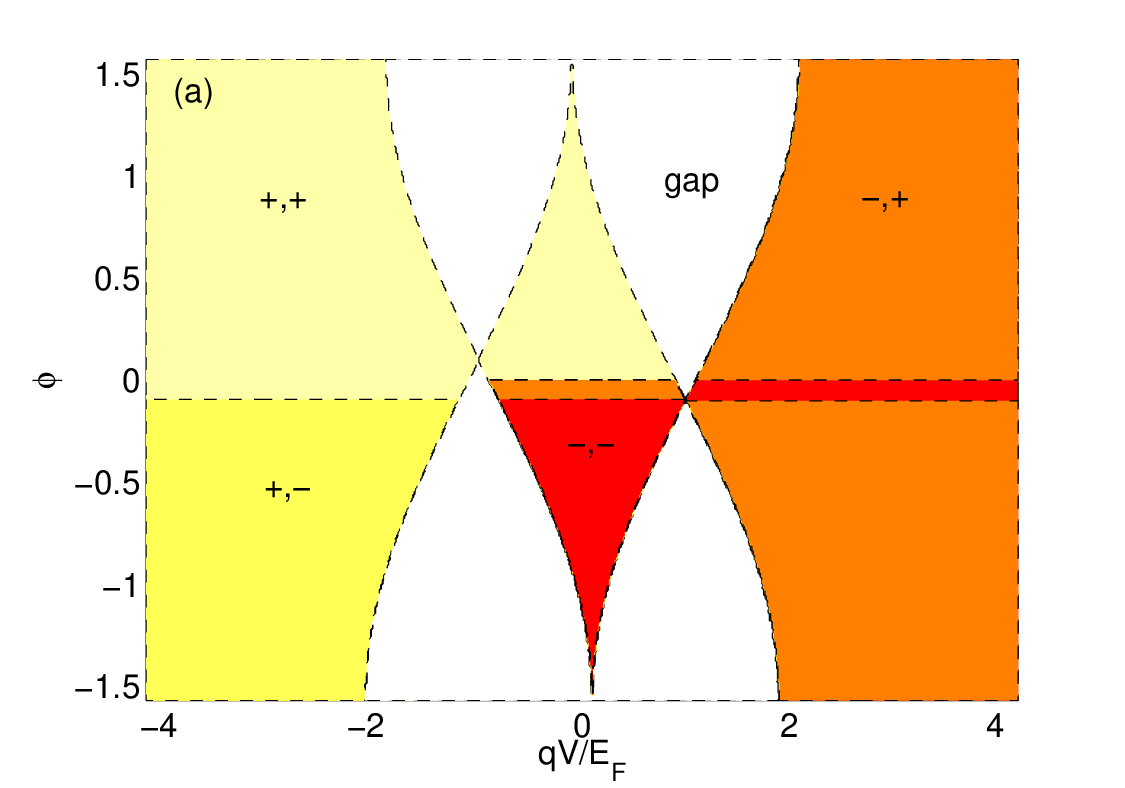} }}
\centerline{{\epsfxsize 7.5cm \epsffile{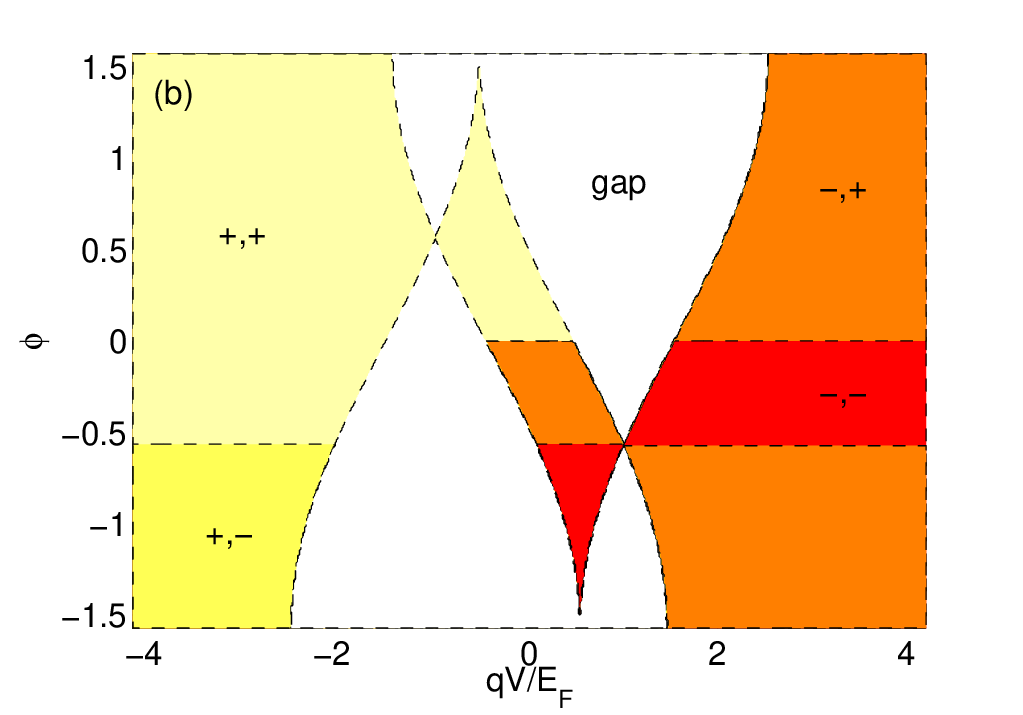}}}
\caption{{ {\it (color online)} Refraction map 
for a double EMVP barrier of field strength (a) $0.1$ Tesla and (b) $3$ Tesla. 
The colouring is for easy identifying sign combinations. 
The first sign is $sgn(\sin |\theta_{1}|)$ and the second sign is $sgn(\sin |\theta_{2}|)$. }}
\label{refrac}
\end{figure}

All these effects  are shown as a refraction map in Fig.\ref{refrac}, which is a combination of  $\sin \theta_{1,2}$ given by Eqns.(\ref{theta1}) and (\ref{theta2}) over a range of $V$.   The white region corresponds to where one of $\sin \theta_{1,2}$ becomes imaginary and as a result the electron wavevector gets totally reflected either from region I  or from region II. The figure shows that both $|$rhs(Eq.(\ref{theta1}))$|$ and$|$rhs(Eq.(\ref{theta2}))$|$ cannot be simultaneously $>1$. The colored portion belongs to where $\theta_{1,2}$ are real, and for the corresponding values  $V$ and $\phi$ there will be transmission.  Thus, the transmission probability greatly increases with the increased value of $|V|$.  

The plots also clearly show that over a range of $\phi$ 
the angles $\theta_{1,2}$ in one of regions I or II  which was imaginary  for a given $V$ becomes real upon increasing $|V|$. However, the sign of $\theta$ become opposite to that of $\phi$ implying negative refraction. This phenomenon explains Klein tunnelling for high $V$ using the language of optics \cite{Falko}. Various sectors of the refraction map also reflect the symmetry properties of such barriers as one makes the transformation $V \rightarrow -V$ and $\phi \rightarrow -\phi$ in Eqs. (\ref{theta1}) and (\ref{theta2}).  

To calculate the transmission, we use a transfer matrix approach and define the following matrices:
\bea M_A = \begin{bmatrix}A & 0 \\ 0 & A^*\end{bmatrix},A=e^{-ik_x}
M_{\theta_{1,2}} = \begin{bmatrix}1 & 1 \\ e^{i\theta_{1,2}} & -e^{-i\theta_{1,2}}\end{bmatrix} \\
M_{s,s_{1,2}} = \begin{bmatrix} 1 & 0 \\ 0 & s,s_{1,2} \end{bmatrix},
 M_{B_{1,2}} = \begin{bmatrix}B_{1,2} & 0 \\ 0 & B_{1,2}^*\end{bmatrix},
 B_{1,2}= e^{-iq_{1x,2x}d} \nonumber \\
& & \label{definematrix}\eea
The continuity of the wave functions at the interfaces 
$x=-d,0,d$ respectively yield 
\bea M_s M_{\phi}M_{A}\begin{bmatrix} 1 \\ r \end{bmatrix} &=&  M_{s_1}M_{\theta_1}M_{B_1}
\begin{bmatrix} a \\ b\end{bmatrix} \nonumber \\
M_{s1}M_{\theta_1}\begin{bmatrix} a \\ b \end{bmatrix} & = & M_{s_2}M_{\theta_2}\begin{bmatrix} c \\ d 
\end{bmatrix} \nonumber \\  
M_{s_2}M_{\theta_2}M_{B_2}^{*} \begin{bmatrix} c \\ d \end{bmatrix}
& = & M_s M_{\phi}M_{A}^{*} \begin{bmatrix} t \\ 0  \end{bmatrix} \eea
The above set of equations can be combined to eliminate $a,b,c,d$ and to obtain transmission coefficient $t$ and reflection coefficient $r$ as 
\beq \begin{bmatrix} 1 \\ r \end{bmatrix} = M_{A}^{*}(M_s M_{\phi})^{-1}
T_{EMVP}M_{s'}M_{\phi}M_{A}^{*} \begin{bmatrix}t \\ 0 \end{bmatrix} \label{demvp}\eeq 
Here, $T_{EMVP}$ is the transfer matrix through such double barrier and is given by
\bea T_{EMVP} & = & M_{s_1}M_{\theta_1}M_{B_1}(M_{s_1}M_{\theta_1})^{-1}M_{s_2} \nonumber \\
              &    &M_{\theta_2}M_{B_2}(M_{s_2}M_{\theta_2})^{-1} \label{Tdemvp}\eea

\begin{figure*}[ht]
\centerline{{ \epsfxsize 15cm \epsffile{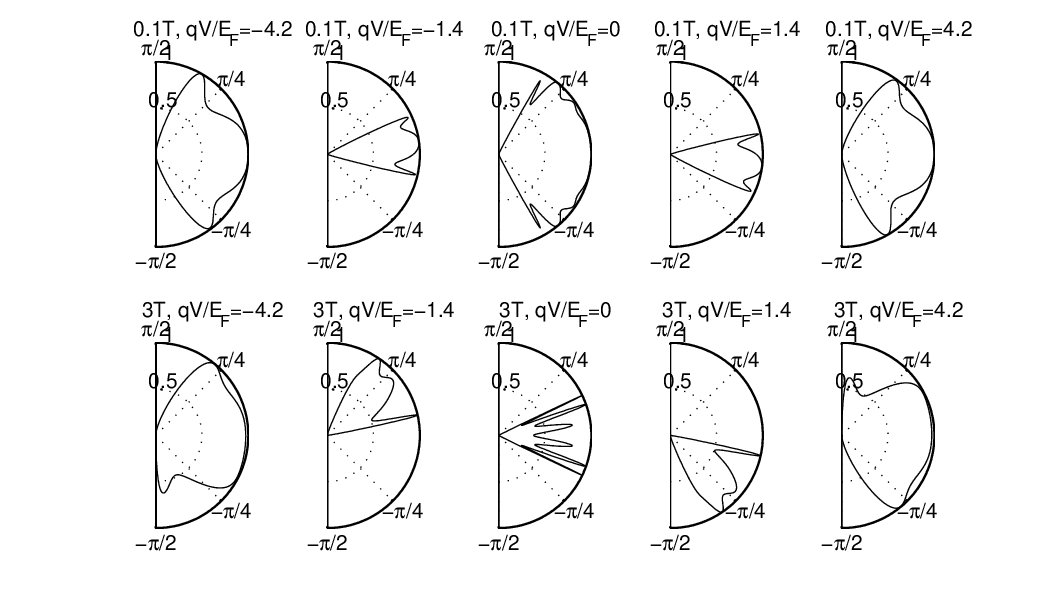}}}
\caption{{Transmittance $T$ through a DEMVP barrier at $0.1$T (upper row)  and $3$T (lower row) at different incidence angles $\phi$.}}
\label{demvpT}
\end{figure*}

Finally, the transmittance $T=t \cdot t^{*}$ is plotted in Fig.\ref{demvpT}. The central column plots the case of zero $V$, where  the DEMVP barrier is equivalent to the DMVP barrier \cite{SGMS}. 

\begin{figure}[ht]
\begin{psfrags}%
\psfragscanon%
%
\psfrag{s05}[t][t]{\fontsize{10}{15}\fontseries{m}\mathversion{normal}\fontshape{n}\selectfont \color[rgb]{0,0,0}\setlength{\tabcolsep}{0pt}\begin{tabular}{c}$qV/E_F$\end{tabular}}%
\psfrag{s06}[b][b]{\fontsize{10}{15}\fontseries{m}\mathversion{normal}\fontshape{n}\selectfont \color[rgb]{0,0,0}\setlength{\tabcolsep}{0pt}\begin{tabular}{c}$\langle T(B,V)\rangle $ (arb. units)\end{tabular}}%
\psfrag{s10}[][]{\fontsize{10}{15}\fontseries{m}\mathversion{normal}\fontshape{n}\selectfont \color[rgb]{0,0,0}\setlength{\tabcolsep}{0pt}\begin{tabular}{c} \end{tabular}}%
\psfrag{s11}[][]{\fontsize{10}{15}\fontseries{m}\mathversion{normal}\fontshape{n}\selectfont \color[rgb]{0,0,0}\setlength{\tabcolsep}{0pt}\begin{tabular}{c} \end{tabular}}%
\psfrag{s12}[l][l]{\fontsize{10}{15}\fontseries{m}\mathversion{normal}\fontshape{n}\selectfont \color[rgb]{0,0,0}$0.1T$}%
\psfrag{s13}[l][l]{\fontsize{10}{15}\fontseries{m}\mathversion{normal}\fontshape{n}\selectfont \color[rgb]{0,0,0}$3.0T$}%
\psfrag{s14}[l][l]{\fontsize{10}{15}\fontseries{m}\mathversion{normal}\fontshape{n}\selectfont \color[rgb]{0,0,0}$1.0T$}%
\psfrag{s15}[l][l]{\fontsize{10}{15}\fontseries{m}\mathversion{normal}\fontshape{n}\selectfont \color[rgb]{0,0,0}$0.1T$}%
%
\fontsize{10}{15}\fontseries{m}\mathversion{normal}%
\fontshape{n}\selectfont%
%
\psfrag{x01}[t][t]{$-4$}%
\psfrag{x02}[t][t]{$-2$}%
\psfrag{x03}[t][t]{$0$}%
\psfrag{x04}[t][t]{$2$}%
\psfrag{x05}[t][t]{$4$}%
%
\psfrag{v01}[r][r]{$0$}%
\psfrag{v02}[r][r]{$0.1$}%
\psfrag{v03}[r][r]{$0.2$}%
\psfrag{v04}[r][r]{$0.3$}%
\psfrag{v05}[r][r]{$0.4$}%
\psfrag{v06}[r][r]{$0.5$}%
\psfrag{v07}[r][r]{$0.6$}%
\psfrag{v08}[r][r]{$0.7$}%
\psfrag{v09}[r][r]{$0.8$}%
\psfrag{v10}[r][r]{$0.9$}%
\psfrag{v11}[r][r]{$1$}%
%
\resizebox{7cm}{!}{\includegraphics{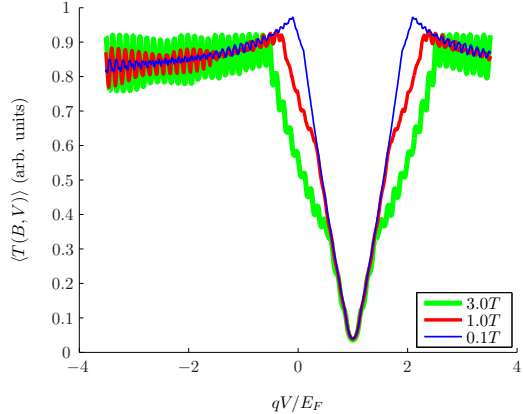}}%
\end{psfrags}%

\caption{{$\langle T(B,V)\rangle$ vs. $V$ through a single EMVP barrier for different B fields. The $x$-axis corresponds to the strength of the potential barrier $\frac{qV}{E_F}$ (dimensionless). }}
\label{singleJ}
\end{figure}

At small values of $V$ the transmission gets asymmetric, and this asymmetry is higher for larger $B$ as also observed earlier for single EMVP barriers. In regions I and II,  the difference in transmission becomes more asymmetric due to different $S_F^{1,2}$. In comparison, with $V=0$ for the DMVP barrier \cite{SGMS}, the refraction properties of right and left side of the barrier compensate each other, with the resultant transmission becoming symmetric.

At very high $V$, the transmission characteristic
is dominated by the electric field. Finite transmission takes place at all $\phi$, and this can also be seen from the corresponding refraction map in 
Fig.\ref{refrac}. As a result, at higher $|V|$, the transmission is less asymmetric (extreme right and left columns). 

\subsection{Effect of various EMVP Barriers on Transport}

\begin{figure}[ht]
\begin{psfrags}%
\psfragscanon%
%
\psfrag{s05}[t][t]{\fontsize{10}{15}\fontseries{m}\mathversion{normal}\fontshape{n}\selectfont \color[rgb]{0,0,0}\setlength{\tabcolsep}{0pt}\begin{tabular}{c}$qV/E_F$\end{tabular}}%
\psfrag{s06}[b][b]{\fontsize{10}{15}\fontseries{m}\mathversion{normal}\fontshape{n}\selectfont \color[rgb]{0,0,0}\setlength{\tabcolsep}{0pt}\begin{tabular}{c}$\langle T(B,V)\rangle$ (arb. units)\end{tabular}}%
\psfrag{s10}[][]{\fontsize{10}{15}\fontseries{m}\mathversion{normal}\fontshape{n}\selectfont \color[rgb]{0,0,0}\setlength{\tabcolsep}{0pt}\begin{tabular}{c} \end{tabular}}%
\psfrag{s11}[][]{\fontsize{10}{15}\fontseries{m}\mathversion{normal}\fontshape{n}\selectfont \color[rgb]{0,0,0}\setlength{\tabcolsep}{0pt}\begin{tabular}{c} \end{tabular}}%
\psfrag{s12}[l][l]{\fontsize{10}{15}\fontseries{m}\mathversion{normal}\fontshape{n}\selectfont \color[rgb]{0,0,0}$0.1T$}%
\psfrag{s13}[l][l]{\fontsize{10}{15}\fontseries{m}\mathversion{normal}\fontshape{n}\selectfont \color[rgb]{0,0,0}$3.0T$}%
\psfrag{s14}[l][l]{\fontsize{10}{15}\fontseries{m}\mathversion{normal}\fontshape{n}\selectfont \color[rgb]{0,0,0}$1.0T$}%
\psfrag{s15}[l][l]{\fontsize{10}{15}\fontseries{m}\mathversion{normal}\fontshape{n}\selectfont \color[rgb]{0,0,0}$0.1T$}%
%
\fontsize{10}{15}\fontseries{m}\mathversion{normal}%
\fontshape{n}\selectfont%
%
\psfrag{x01}[t][t]{$-3$}%
\psfrag{x02}[t][t]{$-2$}%
\psfrag{x03}[t][t]{$-1$}%
\psfrag{x04}[t][t]{$0$}%
\psfrag{x05}[t][t]{$1$}%
\psfrag{x06}[t][t]{$2$}%
\psfrag{x07}[t][t]{$3$}%
%
\psfrag{v01}[r][r]{$0$}%
\psfrag{v02}[r][r]{$0.1$}%
\psfrag{v03}[r][r]{$0.2$}%
\psfrag{v04}[r][r]{$0.3$}%
\psfrag{v05}[r][r]{$0.4$}%
\psfrag{v06}[r][r]{$0.5$}%
\psfrag{v07}[r][r]{$0.6$}%
\psfrag{v08}[r][r]{$0.7$}%
\psfrag{v09}[r][r]{$0.8$}%
\psfrag{v10}[r][r]{$0.9$}%
\psfrag{v11}[r][r]{$1$}%
%
\resizebox{7cm}{!}{\includegraphics{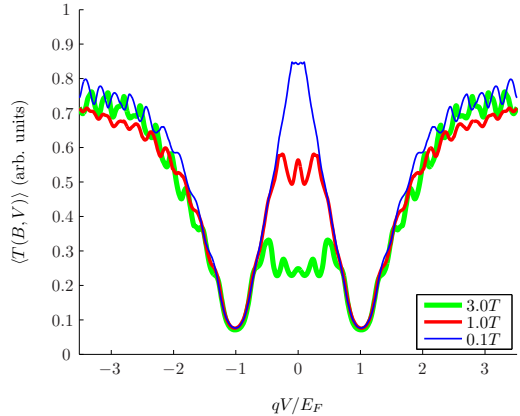}}%
\end{psfrags}%
\caption{{$\langle T(B,V)\rangle$ vs. $V$ through a DEMVP barrier for different B fields. }}
\label{doubleJ}
\end{figure}

To see how the above angle-dependent transmission properties affect electron transport through such barriers, we have also plotted the average transmission through the barrier as a function of the potential $V$ at various strengths of the magnetic barrier. The average transmission at a given  barrier strength $V$ and $B$ is defined as 
\beq  \langle T(B,V) \rangle = v_{F} \int_{-\frac{\pi}{2}}^{\frac{\pi}{2}} d\phi T(\phi, B,V) \cos \phi  \label{current} \eeq
This formula, when generalized to a range of energy levels, leads to the Landauer conductance $G \propto  2 \pi n e^2 \langle T(B,V) \rangle / h$. This has been plotted for three different strengths of magnetic barrier $B$ over a range of  potential barrier strength $V$  for a  SEMVP barrier in Fig. \ref{singleJ} and through a  DEMVP barrier in Fig. \ref{doubleJ}. In case of a SEMVP barrier the transmission shows a minimum as expected  when the point $qV=E_{F}$ is approached. The transmission grows on both side of this singular point and finally at higher $|V|$  oscillates around an average value. For a similar plot of $\langle T(B,V) \rangle$ through the  
DEMVP  barrier given in Fig. \ref{doubleJ} the minimum occurs at two points namely when $qV = \pm E_{F}$. In between this minimum one or more maxima occur depending on the strength of the magnetic barrier $B$. The behavior of the transmission at higher absolute value of the voltage $|V|$ is very similar to that in the case of single EMVP.

Thus, we have shown that adding various forms of electrostatic potential to different magnetic barrier arrangements the symmetry and magnitude of transmission through them can be controlled almost at will. This is not possible for pure magnetic barriers discussed in Ref.\cite{SGMS} and forms another major result in the present work. In Sec.\ref{periodicDEMVP}, we shall extend the transfer matrix approach and show how a sequence of such barriers will modify transmission and the associated band structure. 

Modelling the potential barrier $V$  as a rectangular barrier assumes that the gate voltage-induced doping 
changes abruptly. In reality, however, the doping level continuously changes and thus the edge of the potential 
barrier is actually smooth and not sharp. Thus, the above results may arguably change when a smoother change in the potential is taken into account. This issue has been considered for scalar potential barrier in Ref.\cite{Falko}, where it was found  that a potential which is smooth on the scale of the fermi wave length for small angles of incidence, 
 $T(\phi)=\exp(-\pi k_{F} d \sin^{2} \phi)$ where $d$  is the barrier width.  
 A comparison with Eq.\ref{temvp} shows that the assumption of the rectangular barrier captures the effect of Klein tunneling correctly. However, at other  angles close to the normal incidence it overestimates the transmission.  Transmission through trapezoidal barrier was also analyzed in Ref.\cite{Sonin} which combines the effect of a smooth barrier and a rectangular barrier.


\section{Quantum Goos-H\"anchen Shift in single MVP and EMVP barriers}

\begin{figure}[ht]
\centerline{{ \epsfxsize 6 cm \epsfysize 5.25cm \epsffile{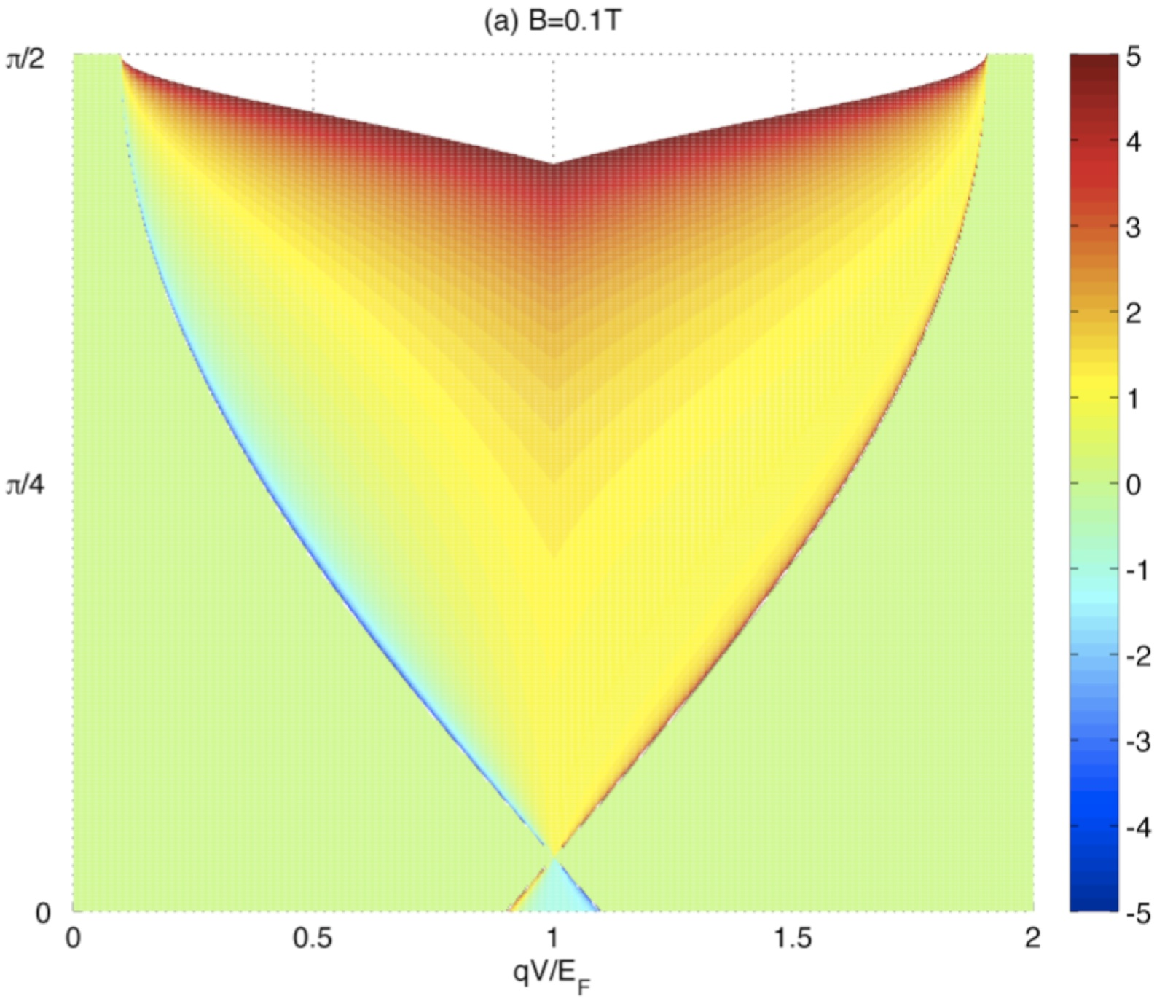}}}
\centerline{{\epsfxsize 6 cm \epsfysize 5.25cm \epsffile{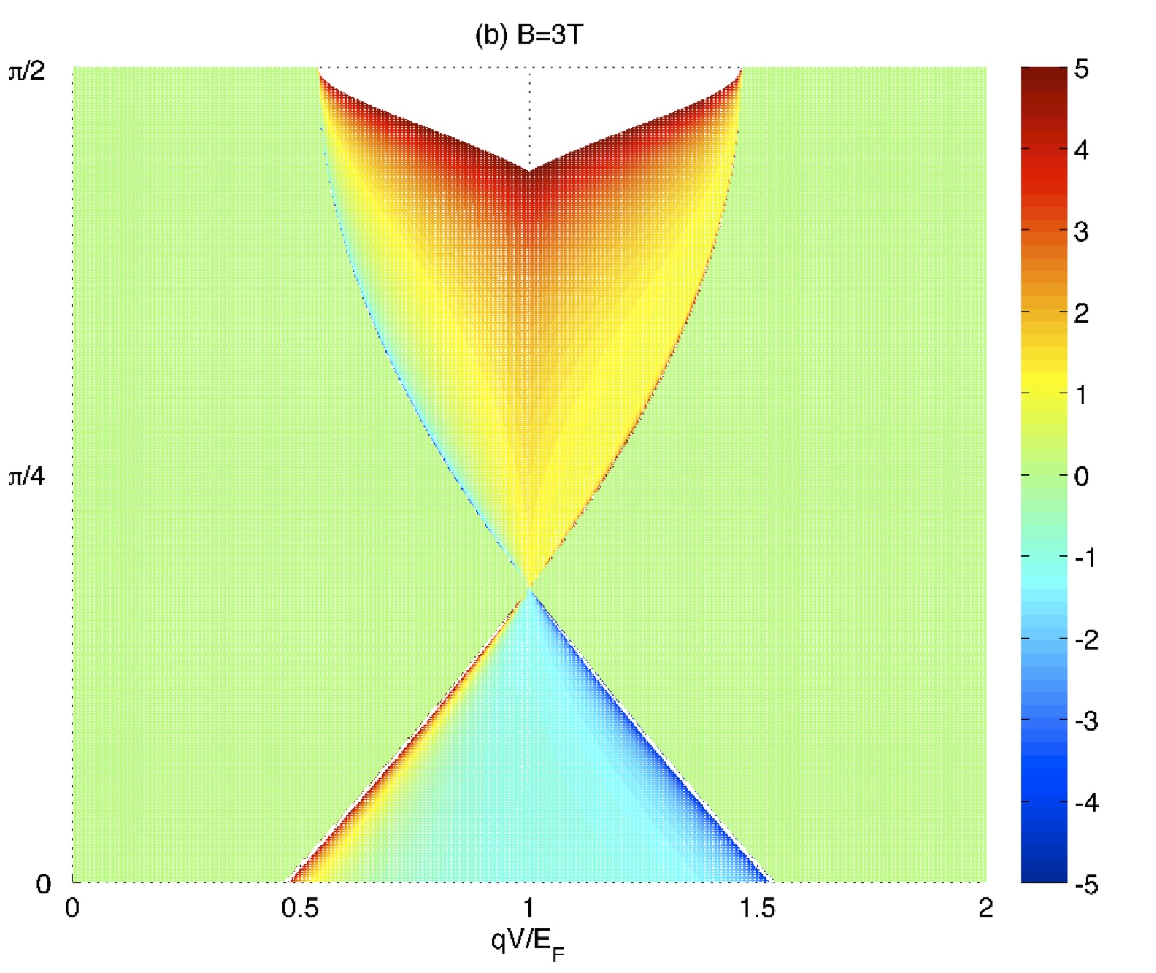}}}
\caption{{{\it (color online)} GH shift for an EMVP barrier at  (a) $0.1$T and (b) $3$T. The $y$-axis corresponds to the incident angle $\phi$. The green regions have propagating solutions with no shift.}}
\label{figGH}
\end{figure}

We now discuss another important optical phenomenon, the Goos-H\"anchen (GH) effect \cite{GH, Newton}. The GH effect describes the shift in a beam of light suffering TIR at an interface along the longitudinal direction ($y$-axis in our discussion). It has been known since the time of Newton \cite{Newton} and was first experimentally measured by Goos and H\"anchen \cite{GH}.  The shift is detectable since the extent of a real beam is always finite. The shift occurs as the totally reflected ray undergoes a phase shift with respect to the incident beam. The shift reverses sign if the second medium behaves like a metamaterial with negative refraction\cite{Berman,Shadridov}.  Such a lateral shift for totally as well as partially reflected  electron waves can also occur for non-relativistic electrons passing through a semiconductor barrier \cite{chen1}, magneto-electric semiconductor nanostructure \cite{chen2}. 

Recently, it has been shown  that ballistic electrons passing through a p-n 
interface in graphene \cite{BeenakkerGH} also suffer a GH shift, which changes sign at certain angles of incidence. We extend this analysis to include the effect of magnetic barriers. We calculate the GH shift using the procedure given in Ref.\cite{BeenakkerGH} for a DEMVP barrier and several such MVP and EMVP barriers as well. 

\begin{figure}[ht]
\centerline{{ \epsfxsize 8cm \epsffile{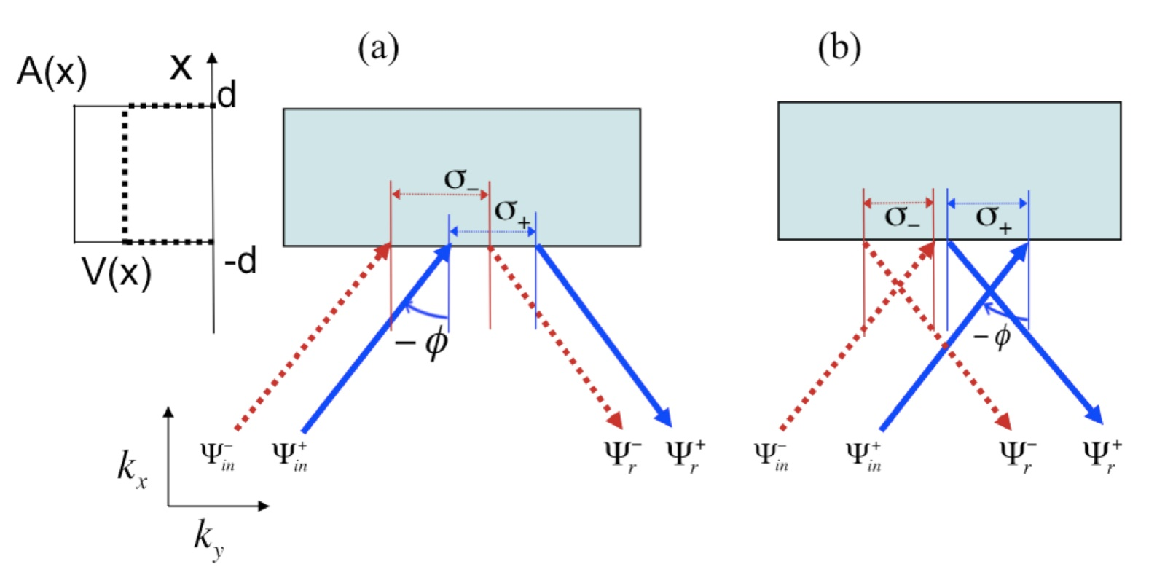} }}
\caption{{ {\it (color online)} GH shift for an EMVP barrier. The solid line corresponds to the upper component of the pseudospinor and the dotted line to the lower component. The GH shift can be either (a) positive or (b) negative.}}
\label{GHray}
\end{figure}

We consider the following wavepacket (beam) of electrons impinging on an MVP or EMVP barrier at $E_F$:  
\beq \Psi_{in}(x,y) = \int_{-\infty}^{\infty} dk_{y} f(k_y-\bar{k}) e^{ik_y y + i k_{x}(k_y) x} \begin{bmatrix} 1 \\ e^{i \phi(k_{y}}) \end{bmatrix} 
\label{profilek} \eeq 
The envelope function ensures the  wavepacket is of finite size along the $y$-direction and is sharply peaked at $k_y=\bar{k}$. 
Thus, 
$\bar{k} \in ( 0,k_F)$ and the angle of incidence $\phi(\bar{k}_y) \in (0, \frac{\pi}{2})$. This fact is represented by writing $k_x$ as well as $\phi$ both as function of $k_y$ in  Eq.(\ref{profilek}). We take a gaussian envelope such that 
\beq  f(k_y - \bar{k}) = \exp[ -\frac{(k_y -\bar{k})^2}{2\Delta_{k}^2}] \label{profile} \eeq 
When $\Delta_{k} \ll k_{F}$, we can approximate the $k_y$-dependent terms by a Taylor expansion around $\bar{k}$ and retaining only the first order term to get 
\beq \phi(k_{y}) \approx \phi(\bar{k}) + \frac {\partial \phi}{\partial k_{y}}|_{\bar{k}}(k_y-\bar{k});~ k_x(k_{y}) \approx k_{x}(\bar{k}) + 
\frac {\partial k_{x}}{\partial k_{y}}|_{\bar{k}}(k_y-\bar{k}) \eeq 
Substituting in Eq.(\ref{profilek}) and integrating, we obtain 
\beq \Psi_{in}  =  \sqrt{2 \pi \Delta_k^2} e^{i [\bar{k} y + k_x(\bar{k}) x]}\begin{bmatrix}
e^{-\frac{\Delta_k^2}{2}[y -\bar{y}_{+}^{in}]^2} \\ e^{-\frac{\Delta_k^2}{2}[y -\bar{y}_{-}^{in}]^2}e^{i \phi(\bar{k})} \end{bmatrix}, \eeq 
where 
\beq \bar{y}_{+}^{in} = -k'_x(\bar{k})x, \bar{y}_{-}^{in}= -k'_x(\bar{k})x - \phi'(\bar{k}) \eeq  
Thus, upper and lower components of the spinorial wave function are localized at separate points along the $y$-axis.  

The reflected wavepacket can also be written in an analogous way by making the transformation $k_x$ to $-k_x$ and $\phi$ to $\pi - \phi$ as 
well as multiplying the reflection amplitude $r(k_y)=|r(k_y)|e^{i \phi_{r}(k_y)}$. 
The reflected wave is then 
\bea \Psi_{r}(x,y) & = & \int_{-\infty}^{\infty} dk_{y} f(k_y-\bar{k}) e^{ik_y y -i k_{x}(k_y) x} \nonumber \\
&  & r(k_y)\begin{bmatrix} 1 \\ -se^{-i \phi(k_{y}}) \end{bmatrix} 
\label{profilekr} \eea 
Here, again $s=sgn(E_F-V)$ and is $1$ for an MVP barrier.
The spatial profile of the reflected wave can be again obtained by first 
expanding all $k_y$ dependent quantities around $\bar{k}$ and retaining only the first order terms and then integrating in Eq.(\ref{profilekr}). This leads to 
\bea \Psi_{r} & = & \sqrt{2 \pi \Delta_k^2} e^{i [\bar{k} y - k_x(\bar{k}) x]} \nonumber \\
& & |r(\bar{k})|\begin{bmatrix}
e^{-\frac{\Delta_k^2}{2}[y -\bar{y}_{+}^{r}]^2} \\ -se^{-\frac{\Delta_k^2}{2}[y -\bar{y}_{-}^{r}]^2}e^{-i [\phi(\bar{k})-\phi_{r}'(\bar{k})]} \end{bmatrix} \eea                                                  Here, $\bar{y}_{+}^{r}$ and $\bar{y}_{-}^{r}$ are given by \beq \bar{y}_{+}^{r} = -\phi'_{r}(\bar{k})+k'_x(\bar{k})x, \bar{y}_{-}^{r}= -\phi'_{r}(\bar{k}) +k'_x(\bar{k})x +\phi'(\bar{k}) \eeq 
The above expression shows that the upper as well as lower components get shifted because of  the phase factor. The GH shifts of the upper and lower components are respectively given by 
\bea \sigma_{+} & = & \bar{y}_{+}^{r} - \bar{y}_{+}^{in} = -\phi_{r}'(\bar{k}) + 2k_{x}'(\bar{k})x \nonumber \\
 \sigma_{-} & = & \bar{y}_{-}^{r} -\bar{y}_{-}^{in} = 2 \phi'(\bar{k}) -\phi_{r}'(\bar{k}) + 2k_{x}'(\bar{k})x \eea  
Thus, the average shift for an MVP or EMVP barrier  is 
\beq \sigma = \frac{1}{2} (\sigma_{+} + \sigma_{-}) = \phi'(\bar{k})-\phi_{r}'(\bar{k}) + 2k_{x}'(\bar{k})x \label{GHshift} \eeq 
The situation is depicted schematically in Fig. 
\ref{GHray}. The last term in the above expression is a coordinate dependent quantity and will get an equal and opposite contribution from the $-\phi_{r}'(\bar{k})$ term. The resultant $\sigma$ will thus be independent of the choice of the coordinate of the interface from which TIR will take place. Thus, we can calculate the GH shift when the angle of incidence $\phi$ is greater than the critical angle of incidence $\phi_c$.

For the case of either an EMVP barrier or an MVP barrier, the reflection coefficient $r(k_{y})$ can be calculated in the same way as done for the transmission coefficient given in Eq.(\ref{temvp}) by demanding the continuity of  wave functions on both sides of the barrier at $x=-d$, and noting that on one side of the barrier the wave function is evanescent. Using Eqs.(\ref{particle1}) and (\ref{hole1}), such a wave function can be written as 
\beq
\phi_1^{GH} = \left\{
\begin{array}{rl}
e^{ik_x x} + r e^{-ik_x x} & \text{if }~~ x < -d\\
a' e^{-\kappa (x+d)}  & \text{if }~~ x >-d  \\
\end{array} \right.
\label{particlegh}\eeq
\beq \phi_2^{GH} = \left\{
\begin{array}{rl}
s[e^{i (k_x x +  \phi)} - r e^{-i(k_x x + \phi)}] & \text{if }~~ x < -d\\
-i \gamma s' a'e^{-\kappa (x+d)}  &   \text{if }~~ x > -d 
\\ 
\end{array} \right . 
\label{holegh}
\eeq
Here $s,s'=sgn(E_F-V)$, and  
\bea \gamma & = & \frac{\kappa + (k_{y} - \frac{1}{\ell_B})}{k_F}, 
(k_{y} - \frac{1}{\ell_B})^2  - \kappa^2  =  k_F^2
~~\text{MVP}  \nonumber \\
      \gamma & = & \frac{\kappa + (k_{y} - \frac{1}{\ell_B})}{k_F'}, 
(k_{y} - \frac{1}{\ell_B})^2  -\kappa^2  =  k_F'^{2}
~~\text{EMVP} \label{evan1} \eea 
where $k_{F}'$ has been defined in the Eq.(\ref{energyemvp}). 
For the wavepacket considered above, these  conditions for evanescent wave may not be satisfied for different $k_y$ that constitute the wavepacket and only a fraction of such 
components may be totally reflected. However, as long as $\Delta_k \ll \bar{k}$ it is reasonable to assume that the entire wavepacket is either partially transmitted or fully reflected. If, on the other hand, one considers  a wave packet which is broader then the above conditions need to be relaxed. 

In view of Eq.(\ref{evan1}), it is instructive to parametrize
$\kappa$, $k_y -\frac{1}{\ell_B}$, for a totally reflected wave packet, in one of the following two alternative ways:  
\bea |k_{y} - \frac{1}{\ell_B}| & = & k_F'(k_F) \cosh \alpha ; \kappa = k_F'(k_F) \sinh \alpha ,~~\text{(E)MVP}\nonumber \\
\text{or}, k_F'(k_F)& = & |k_{y} - \frac{1}{\ell_B}|\sin \beta ; \kappa = |k_y - \frac{1}{\ell_B}| \cos \beta ,~~\text{(E)MVP} \nonumber \\
\text{with}~  \gamma & = & \exp(\alpha)=
\cot (\frac{\beta}{2}) \nonumber 
\eea 
The second parametrization can be heuristically understood as the angle made by the wavevector of the totally reflected wave 
with the surface normal in the first medium. The continuity of the wave function at $x=-d$ gives the reflection coefficient as 
\bea r & = & e^{-ik_x D} [\frac{ie^{i \phi} - ss'\gamma}{ie^{-i \phi} + ss' \gamma} ]=\exp(-ik_x D)\exp(2i\delta) \nonumber \\
\text{with} \tan \delta & = & \tan \phi + ss' \gamma \sec \phi \eea 
This expression is similar to the one derived for scalar electrostatic barrier \cite{BeenakkerGH} with the exception of the prefactor $e^{-i k_x D}$, which appears due to a different choice of origin and does not affect the GH shift in Eq. (\ref{GHshift}) as explained earlier. The reflection coefficient $r$ is expectedly a unimodular complex number with the phase given by 
\beq \phi_{r} = -k_x D + 2 \delta \eeq
The GH shift now can be rewritten as 
\bea \sigma & = & \phi'(\bar{k}) - 2 \delta'(\bar{k}) \label{ghfinal} \\
\text{where} \delta & = & \tan^{-1}(\tan \phi - \Delta) \nonumber \eea 
Using this expression, one can calculate the GH shift for MVP and EMVP barriers in the case of TIR. 

\subsubsection{GH shift for MVP barriers}

For pure MVP barrier, TIR will take place only when $0 > \phi > -\frac{\pi}{2}$
since the wave incident  from the right and left hand side of the surface normal will behave differently \cite{SGMS}. Then, the critical angle for TIR is 
\beq |\phi_{c}|=\sin^{-1}[1-\frac{1}{k_F \ell_B}] \nonumber \eeq 
Thus, there will be a finite GH shift for $ -\phi_{c} \ge \phi \ge -\frac{\pi}{2}$. 

\subsubsection{GH shift for EMVP barriers}

For an EMVP barrier, TIR occurs for electrons incident from both sides of the surface normal but at different critical angles. For a given $\phi$, it is possible to change $V$ adiabatically and get the electron wave reflected over a range of $V$ satisfying $|\sin \theta| > 1$. We can keep $\phi$ fixed and increase $V$. At $V=V_{c1}$, TIR occurs when 
\bea \sin |\theta| & = & 1, \nonumber \\
\Rightarrow \frac{V_{c1}}{E} & = & 1-[\sin|\phi| - \frac{1}{k_F \ell_B}], 0 < \phi < \frac{\pi}{2} \nonumber \\
& = & 1-[\sin|\phi|+\frac{1}{k_F \ell_B}], 0 > \phi >   \frac{\pi}{2} \eea 

Upon further increasing $V$, the electron wave remains totally  
reflected till $V$ reaches the second critical value $V_{c2}$ such that  
\bea 
\sin |\theta| & = & -1 \nonumber \\
\frac{V_{c2}}{E} & = & 1 + [\sin |\phi| - \frac{1}{k_F \ell_B}], 0 \le \phi \le \frac{\pi}{2} \nonumber \\  
& = & 1 + [\sin |\phi| + \frac{1}{k_F \ell_B}], 0 \ge  \phi \ge \frac{\pi}{2}
\eea 

TIR occurs in the range $V \in [V_{c1} , V_{c2}]$. In Fig. \ref{figGH} 
is plotted the GH shift over the entire range of $V$ and $\phi$.
At all other regions in the $\phi - V$ plane the GH shift is set to $0$ (green). The point $V=E_F$ where the scale factor diverges lies within this range. At this singular point, the GH shift should be calculated by taking into account the special nature of the solutions given by Eq. \ref{zeromodes}.  At all other $V$, the GH shift can be calculation from Eq.(\ref{ghfinal}). The explicit expression for the GH shift for an EMVP barrier in dimensionless form is 
\beq 
\sigma k_F  =  \frac{1 - 2 \frac{ k_F}{k_F'}\cos^2 \delta(1+ \tan \phi \tan \delta + ss' \frac{\gamma^2 +1}{\gamma^2 -1})} {\cos \phi}
\label{fullGH} \eeq
The $ \frac{\gamma^2 +1}{\gamma^2 -1}$ term in the numerator which is  $\coth \alpha$ or $\sec \beta$ 
diverges at the critical angle for TIR  and leads to a divergent GH shift. This is because just at the critical angle the wavevector lies in the interface of the two regions.
As a result in  Fig. \ref{figGH}, the border TIR region of finite GH shift shows the highest GH shift. 

The lower and upper parts of the curve again correspond to $\phi <\phi_s$ and  $\phi > \phi_s$ as seen earlier in Section \ref{emvp1}. The left and right boundaries correspond to $V < E_F$ and $V > E_F$. In the lower part, the GH shift is mostly negative with the left and right boundaries having positive and negative refraction respectively.

Comparing Eq.(\ref{fullGH}) with Eq.11 of  Ref.\cite{BeenakkerGH}, we see that $\sigma(\phi) \ne - \sigma(-\phi)$ in our case. This is due to the non-specular nature of electron refraction at an EMVP barrier. This is an important difference 
in the quantum GH effect that occurs upon TIR by  an EMVP barrier as compared to TIR by a purely electrostatic barrier. 

The GH shift  can be calculated similarly at a DEMVP barrier, but the wave could suffer TIR either at the first or the second interface. One needs to ascertain first which interface is causing TIR. Once that is determined, the GH shift can be calculated as described for the single EMVP barrier. After TIR, the wave will be GH shifted and not propagate further. For an array of such barriers, the fraction of the incident electrons suffering GH shift will be enhanced. Such a quantum GH effect can lead to interesting devices in the coherent transport regime \cite{Yellin}. 

As discussed in Ref.\cite{BeenakkerGH}, the GH shift affects the quantum transport in a pronounced manner. Here we will describe this aspect very briefly in a qualitative way. Upon multiple reflections from the interface, the GH shift gets accumulated. 
The shift causes a change in the velocity component that is parallel to the interface. This  is different for the up and down pseudospin components and in turn lifts the degeneracy  in the velocity component of the transport electrons by changing their dispersion. 

According to Landauer formula, in the ballistic regime the conductance is obtained by summing over transmittance $T$ from all channels due to spin and valley degeneracy. Once the GH shift is taken into account for each such channel,  the upper and lower components of the pseudopsinors having two different velocities along the interfaces will contribute two more additional channels for such transport. As a result, the conductance will increase at the appropriate width of such a barrier . Similar consideration 
for electron transport in the presence of GH shift  is valid for  MVP and EMVP barriers as well.


\begin{figure*}[ht]
\centerline{{\epsfxsize 14cm  \epsffile{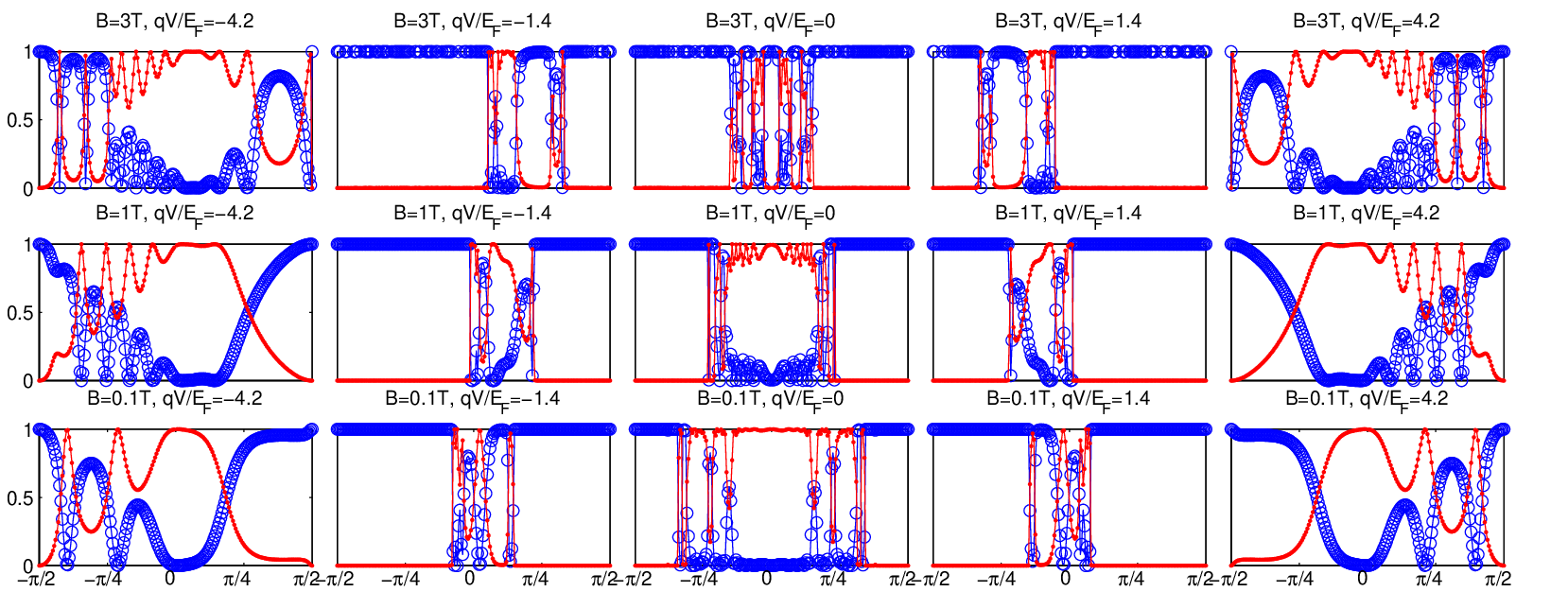}}}
\centerline{{ \epsfxsize 14cm  \epsffile{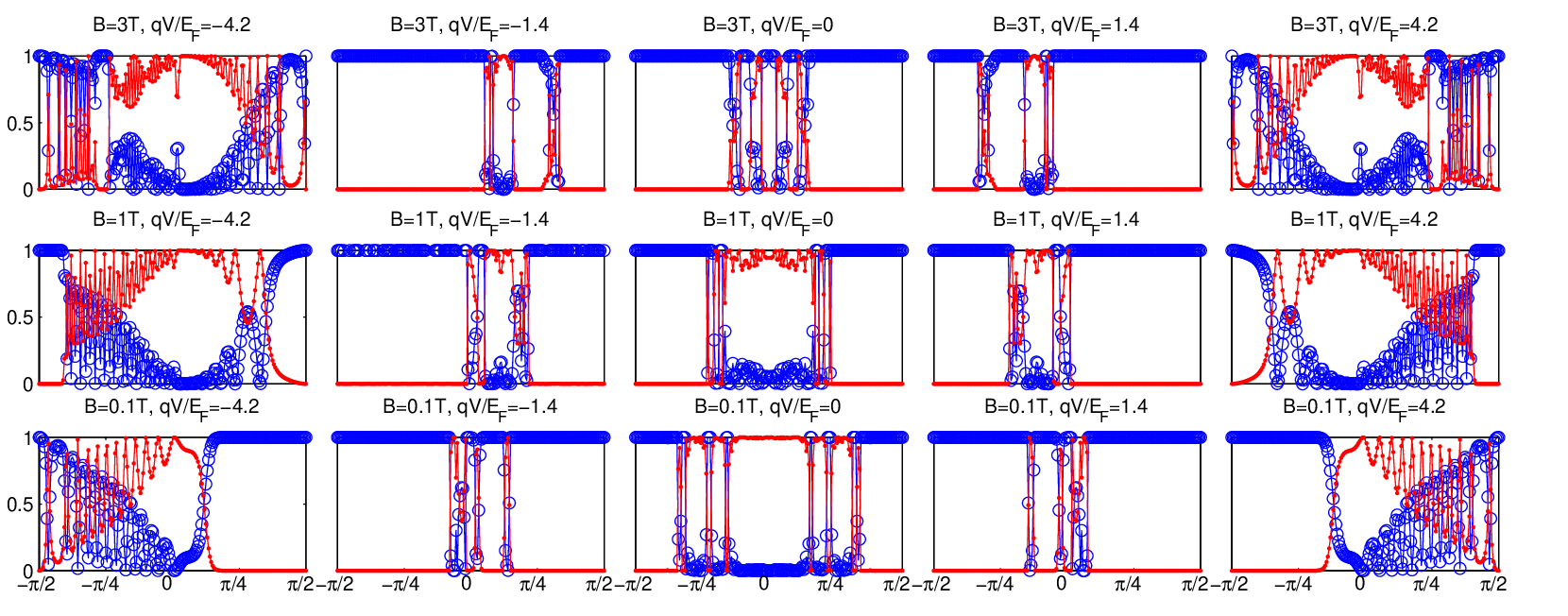}}}
\caption{{\it (color online)} Transmittance (red dots) and Reflectance (blue circles) through a finite sequence of $5$ (upper three rows) and $25$ (lower three rows) DEMVP barriers. On the $x$-axis is plotted the incident angle $\phi$.}
\label{fig6}
\end{figure*}

\section{Periodic lattices of DEMVP barriers}\label{periodicDEMVP}

We now look at  periodic arrangements of DEMVP barriers considered individually in Section \ref{doubleEMVP}. We first consider a finite number of DEMVP barriers with suitable  boundary conditions and then extend the results to an infinite periodic lattice.

\subsection{A finite lattice of DEMVP barriers}

A Bragg reflector is created with  $N$ double 
EMVP barriers of Fig.\ref{ferromag}(b) placed symmetrically about the origin with the magnetic field given by\cite{SGMS}
\bea
\bs{B} = B_z(x)\hat{z} &=& B\ell_B[\delta(x+nd)+ \delta(x-nd) \nonumber \\
 & & + \sum_{p=1-n}^{n-1}(-1)^{p+n}2B\delta(x-p d))]\hat{z}   \eea
The wave function solutions in various regions are linear combinations of right and left moving waves similar to Eqs. (\ref{particle2}) and (\ref{hole2}). The transmission coefficient can be calculated by using the continuity of the wavefunction in the interface of two such barriers. 
With $T_\text{EMVP}$ defined as in Eq.(\ref{Tdemvp}), 
the transfer matrix for $n$ DEMVP barriers is
\beq \begin{bmatrix}1  & r \end{bmatrix}^{T}  = (M_{A}^{-1})^{n}(M_s M_{\phi})^{-1}
T_{EMVP}^{N}M_{s'}M_{\phi}(M_{A}^{*})^{N} 
\begin{bmatrix}t & 0\end{bmatrix}^{T}\eeq 

Plots for $N=5$ and $N=25$ are given in Fig.\ref{fig6}. 
The case $V=0$ (central column) corresponds to  
 pure double MVP barriers \cite{SGMS} and are shown here for comparison. With increasing $|V|$, reflectance gets reduced. For low $|V|$, reflectance is highly asymmetric due to the asymmetric behaviour of $\sin \theta_1$ and $\sin \theta_2$ with $\phi$ (Fig.\ref{refrac}). For higher $|V|$, full transmission occurs over the entire range of $\phi$. This significantly decreases the reflectance of the Bragg reflector. However, with more DEMVP barriers this decay is slowed down and is preceded by an oscillatory behaviour of reflectance.

\subsection{An infinite periodic lattice of DEMVP barriers}

Though any realistic system is finite, to understand the transport in a sufficiently large sample it is instructive to ignore boundary effects and  assume the double EMVP barrier structure of Eq.(\ref{split-gate}) can be repeated infinitely along the $x$-axis. 
Since changing or reversing $V$ can locally convert a charge-neutral region into a p-n or n-p junction, such a periodic barrier can also be thought of as a semiconductor heterostructure. It should be emphasized that this can lead to new device structures due to combined effect of the highly inhomogeneous and periodic magnetic fields and controllable voltages. 

Consider each unit cell of size $D=2d$ for the MVP as well as for the electrostatic potential barriers. 
Thus, the n-th cell is given by the region between  $(n-1)D$  and $nD$. In the $\alpha$-th part of a unit cell, the wavefunction is
\bea \phi_1 & = & a_{n}^{\alpha} e^{i q^{n}_{\alpha x} (x - nD)} + b_{n}^{\alpha}e^{-i q^{n}_{\alpha x} (x - nD)} \\
\phi_2 & = & s_{n}^{\alpha} \left[a_{n}^{\alpha} e^{i [q^{n}_{\alpha x} (x - nD)+ \theta_{\alpha}]} - b_{n}^{\alpha}e^{-i [q^{n}_{\alpha x} (x - nD)+ \theta_{\alpha}] } \right] \nonumber \label{periodicsol} \eea
Here, \bea \alpha  =  1,2; a_{n}^{1}  =  a_n, b_{n}^{1}  =  b_n, & & 
 a_{n}^{2}  =  c_n, b_{n}^{2}  =  d_n; \nonumber \\
s_n^{1,2}=s_{1,2}; s_{1,2}=\text{sgn}(E　\pm V) & & 
q^{n}_{1x,2x} =  q_{1,2} \eea The exponential factor $e^{-nD}$ reveals the existence of 
lattice translational symmetry, which is not present for the isolated EMVP and DEMVP barriers of Section \ref{secIII}.
 The wavevectors $q_{(1,2)}$  are given by  Eqs.(\ref{dem2}) and (\ref{dem3}) and are different from pure magnetic barriers.  Also, $s_{1,2}$ can have same or opposite sign depending on $V$ and $E$, whereas for pure MVP barriers they have  the same sign. These differences  from pure magnetic barriers of Ref.\cite{SGMS} are reflected strongly in the band structure and are another major aspect of this work. 

The continuity of the wavefunction at the first interface at $x=(n-1)D$ gives
\bea 
&  & \begin{bmatrix}1 & 1 \\ s_{2}e^{i \theta_2} & - s_2e^{-i \theta_2} \end{bmatrix} \begin{bmatrix} c_{n-1} \\ d_{n-1} \end{bmatrix} 
   \nonumber \\
&  & = \begin{bmatrix} e^{-i q_1 D}  & b_n e^{i q_1 D}  \\
  s_1 e^{i[q_1 D - \theta_1]} & - s_1 e^{i[q_1 D 
- \theta_1]}] \end{bmatrix} \begin{bmatrix} a_n \\ b_n \end{bmatrix}  \label{conti1} \eea

Similarly, the continuity at the second interface at $x=(n-1)D+ d$ gives 
\bea & & \begin{bmatrix} e^{-iq_1\frac{D}{2}}  & e^{i q_1 \frac{D}{2}} \\
s_{1}  e^{-i[ q_1 \frac{D}{2}- \theta_1}]  & -s_1 e^{i[ q_1 \frac{D}{2}- \theta_1]}] \end{bmatrix} \begin{bmatrix}a_n \\ b_n \end{bmatrix} \nonumber \\ 
 &  & = 
\begin{bmatrix} e^{-i q_2 \frac{D}{2}} &   e^{i q_1 \frac{D}{2}}  \\ 
s_2[ e^{-i[ q_2 \frac{D}{2}- \theta_2}]  & - s_2e^{i[ q_2 \frac{D}{2}-
\theta_2]}] \end{bmatrix} \begin{bmatrix}c_n \\ d_n \end{bmatrix} \label{conti2} \eea 

Using Eq.(\ref{definematrix}), Eqns.(\ref{conti1}) and (\ref{conti2}) can be rewritten as 
\bea M_{s_{{2},{n-1}}}M_{\theta_2}\begin{bmatrix} c_{n-1} \\
d_{n-1} \end{bmatrix} &= & M_{s_{1,n}}M_{\theta_1} {M_{B_1}}^2\begin{bmatrix} a_{n} \\
b_{n} \end{bmatrix} \nonumber \\
M_{s_{1,n}}M_{\theta_1}M_{B_1}\begin{bmatrix} a_{n} \\ b_{n} \end{bmatrix} & = &
M_{s_{2,n}}{M_{\theta_2}}{M_{B_2}}\begin{bmatrix} c_{n} \\ d_{n} \end{bmatrix} \label{latticebc}\eea

The above two matrix equations can be combined as 
\begin{widetext}
\beq \begin{bmatrix} c_{n-1} \\ d_{n-1} \end{bmatrix}  = 
M_{\theta_2}^{-1}M_{s_2}^{-1}M_{s_1}M_{\theta_1}M_{B_1}M_{\theta_1}^{-1}M_{s_1}^{-1}M_{s2}M_{\theta_2}M_{B_2}\begin{bmatrix} c_{n} \\ d_{n} \end{bmatrix}  =  \begin{bmatrix} K_{11} & K_{12} \\
K_{21} & K_{22} \end{bmatrix} \begin{bmatrix} c_{n} \\ d_{n} 
\end{bmatrix}\label{Blochcond1}\eeq
\end{widetext}
According to Bloch theorem,
\beq \begin{bmatrix} c_{n-1} \\ d_{n-1} \end{bmatrix} = e^{-iK D} \begin{bmatrix} c_{n} \\ d_{n} \end{bmatrix} \label{Blochcond2} \eeq 
The matrix $K_{mat}=\begin{bmatrix} K_{11} & K_{12} \\
K_{21} & K_{22} \end{bmatrix}$ is unimodular.  
From Eqs.(\ref{Blochcond1}) and (\ref{Blochcond2}) we obtain the eigenvalue equation 
\beq 
\begin{bmatrix} K_{11} & K_{12} \\
K_{21} & K_{22} \end{bmatrix} \begin{bmatrix} c_{n} \\ d_{n} \end{bmatrix}
 =e^{-iK D} \begin{bmatrix} c_{n-1} \\ d_{n-1} \end{bmatrix} \eeq
where $K$ is the Bloch momentum. The various terms can be written explicitly as 
\bea K_{11} & = & F^{+}(\theta_2, \theta_1)F^{+}(\theta_1, \theta_2) e^{-i(q_1+q_2)d} \nonumber \\
& & + F^{-}(\theta_2, \theta_1)F^{-*}(\theta_1, \theta_2)e^{i(q_1 - q_2)d}
\nonumber \\
K_{12} &= & F^{+}(\theta_2, \theta_1)F^{-}(\theta_1, \theta_2) e^{-i(q_1-q_2)d} \nonumber \\
&  & + F^{-}(\theta_2, \theta_1)F^{+*}(\theta_1, \theta_2)e^{i(q_1 - q_2)d} \nonumber \\
K_{21}& = & F^{-*}(\theta_2, \theta_1)F^{+}(\theta_1, \theta_2)e^{-i(q_1 + q_2)d} \nonumber \\
&  & + F^{+*}(\theta_2,\theta_1)F^{-*}(\theta_1, \theta_2)e^{i(q_1 -q_2)d} \nonumber
\\
K_{22} & = & F^{+*}(\theta_2,\theta_1)F^{*-}(\theta_1, \theta_2)e^{i(q_1-q_2)d} \nonumber \\
&  & + F^{-*}(\theta_2, \theta_1)F^{+}(\theta_1, \theta_2)e^{-i(q_1 + q_2)d}
\eea  
where 
\beq F^{\pm}(\theta_{k}, \theta_{l})  =  e^{-i \theta_k} \pm s_1s_2
e^{i \theta_l}, \text{for}~ k,l=1,2.\nonumber \eeq 
The complex conjugate eigenvalues $\lambda$ are  given by  
\bea det|K_{mat} - \lambda I| & = & 0  \nonumber \\
\Rightarrow \lambda_{1} + \lambda_{2} & = & \exp(-i K D) + \exp(i K D), 
\eea 
which finally gives 
\beq K(\phi, B)=
\frac{1}{2d} \cos^{-1}[\frac{1}{2}Tr(K_{ij})] \label{eqbandstructure} \eeq
The condition $|\frac{1}{2}Tr(K_{{ij}})| < 1$ corresponds to propagating Bloch waves whereas $|\frac{1}{2}Tr(K_{ij})| > 1$ leads to evanescent Bloch waves that correspond to forbidden zones in the band structure in presence of periodic DEMVP  barriers.
Writing in terms of the wavevectors $q_1,q_2$ and the angles $\theta_1, \theta_2$, the above eigenvalue condition reads 
\bea \cos KD & = & \cos q_1 d \cos q_2 d + \sin q_1 d \sin q_2 d \times \nonumber \\
&  & \big[
\tan \theta_1 \tan \theta_2 + \frac{s_1s_2}{\cos \theta_1 \cos \theta_2}\big]
\label{band1} \eea

\begin{figure*}[ht]
\centerline{{ \epsfxsize 16cm  \epsffile{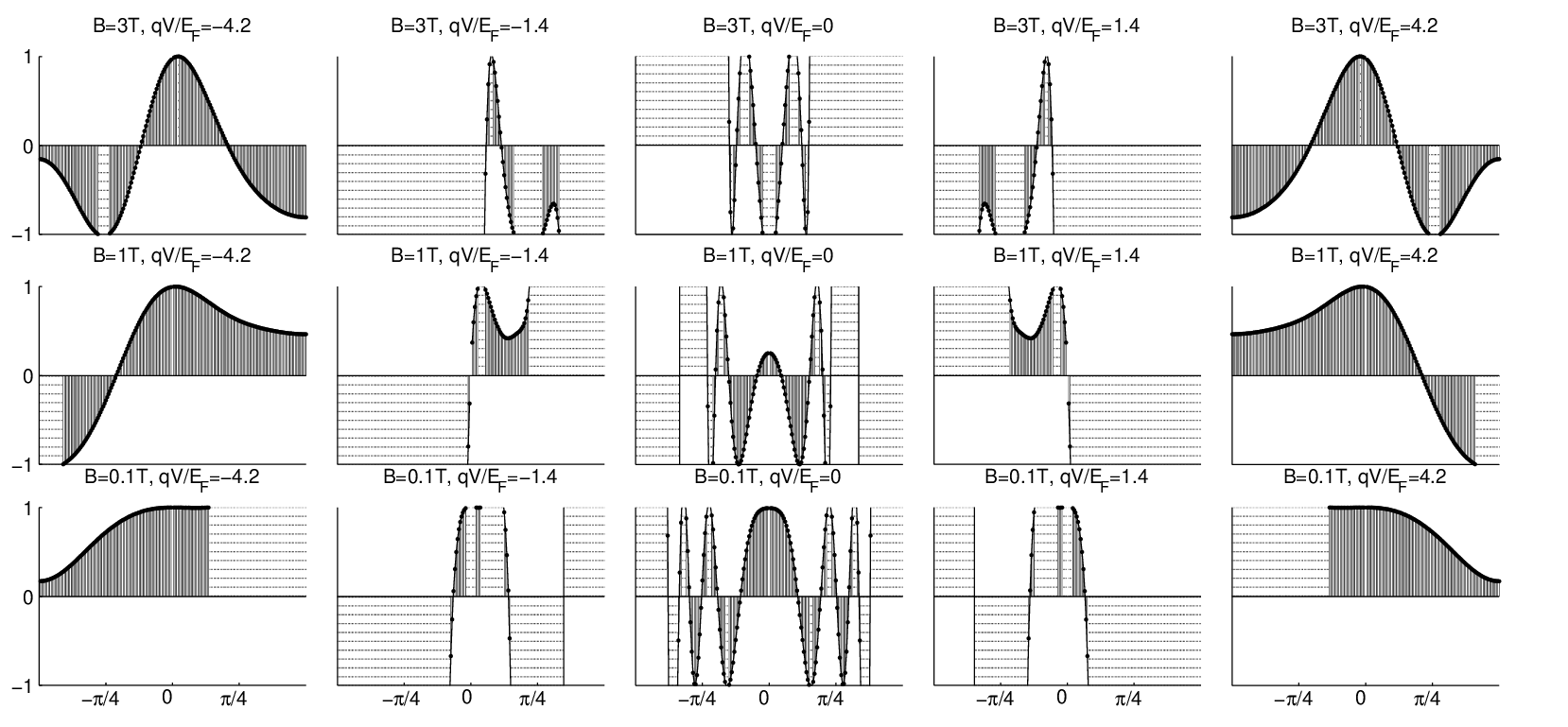}}}
\caption{{Plot of $\cos(KD)$ vs. $\phi$ giving allowed (black) and forbidden (hashed) regions for EMVP barriers of $0.1$, $1$ and $3$ Tesla. }}
\label{emvpband1}
\end{figure*}

The above equation provides the band structure for a periodic DEMVP barrier. This is an extension of the Kronig-Penney (KP) model to two-dimensional massless Dirac fermions. Thus, it is interesting to compare the DEMVP band structure  with other variants of the KP model. 
The original KP model describes Bloch waves in a one-dimensional periodic potential \cite{Kittel}. Several authors have also studied the relativistic KP model \cite{periodic1, periodic2, periodic4, periodic5}, where the motion considered is strictly one-dimensional. The non-relativistic KP model in periodic structures created by MVP barriers has also been studied \cite{MPV94,periodic3}. This second set of studies for non-relativistic electrons has a different set of boundary conditions at the unit cell interfaces since the generic wave equation is of second order. 

The current problem is a relativistic KP model for two dimensional massless Dirac fermions in graphene in the presence of an effectively one-dimensional potential. Previously, such problems have been studied for different types of periodic magnetic 
\cite{SGMS, Martino} as well as electrostatic \cite{periodic6} barriers. The present analysis is new as it combines the effects of electrostatic and magnetic barriers.

In Fig.\ref{emvpband1}, Eq.(\ref{band1}) is used to plot $\cos KD$ versus $\phi$
 for three different $B$ fields to determine the real and imaginary Bloch wave vector $K$. The central column shows  symmetric transmission at zero $V$, which is an infinite series of MVP barriers \cite{SGMS}. The forbidden region is placed symmetric around zero $\phi$ and widens at higher $B$. 


At $\pm qV/E_F=1.4$, the region of allowed propagation shrinks and gets asymmetrically distributed about the $\phi$ axis. At $\pm qV/E_F=1.4$, beyond a certain $\phi$ 
there is no transmission. When $V$ is increased further  to $\pm qV/E_F=4.2$, for one sign of $\phi$ all solutions lead to propagating Bloch waves whereas for the opposite sign of $\phi$ such  propagating solutions exist only over some part.  
Upon reversing $V$, the behaviour gets mirrored about the $\phi$-axis. This asymmetric dependence on $\phi$ as well as reversal upon the change of sign in $V$ persist for other values $B$. When $B$ is increased from $0.1$T to $3$T, more forbidden zones form in between the conducting regions on the $\phi$ axis. These results are consistent with those for a single DEMVP barrier in  Fig.\ref{refrac} and Fig.\ref{demvpT}.

In this regime, $E=\hbar v_{F}|k|$ and  $k_{y}=|k| \sin \phi$. For a given $E$, $|k_{y}| \in \{\frac{E}{\hbar v_{F}},0 \}$ and $|\phi| \in \{ \frac{\pi}{2},0 \}$. Given $\phi$, it is possible to determine using Eq.(\ref{band1}) whether Bloch waves will be propagating or evanescent, the latter case corresponding to the forbidden zone.

\begin{figure*}[ht]
\centerline{{\epsfxsize 16cm  \epsffile{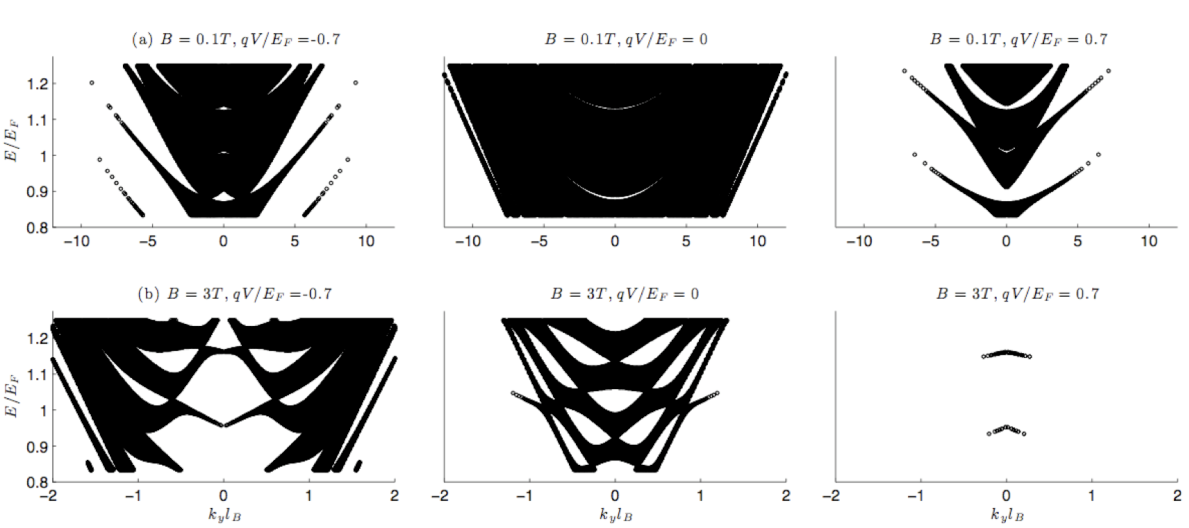}}}
\caption{{Band structure at some characteristic values of $V$ with 
$B$-field strength strengths (a) $3$ Tesla, and (b) $0.1$ Tesla.}}
\label{emvpbandky}
\end{figure*}

In Fig. \ref{emvpbandky} is plotted the band structure of corresponding to the periodic EMVP barriers as a function of magnetic field $B$ and voltage $\frac{qV}{E_{F}}$. Conducting regions are seen over a wide range of $E$ with forbidden regions in between. At $0.1$T, the region near $V=0$ is mostly conducting for different values of energy where a forbidden region starts opening up  both to the left and right of the $V$ axis. The gaps are much wider than their counterparts for pure MVP barriers.  

It is emphasized again that, exactly at the singular point ${qV}=E_F$, the 
method of obtaining band structure using Eq.(\ref{band1}) is not valid since the solutions are always evanescent as explained by Eq.(\ref{zeromodes}). Instead, if we analyze close to ${qV}\rightarrow E_F$, an extended forbidden zone appears. 
To understand more explicitly, the band structure is also plotted at $\frac{qV}{E_F}=0.7$ close to the singular point. 
Here, the conducting regions are intervened by large patches of forbidden zones. In comparison, at $\frac{qV}{E_F}=-0.7$, the behaviour is completely changed and the forbidden zone over the same range of $k_y \ell_B$ shrinks considerably (left column). This can be directly related to the laws of refraction for a DEMVP barrier as in Eqs. (\ref{theta1}) and (\ref{theta2}). 
For higher $V$ where $qV >E_F$, the system is conducting at almost all $E$. This is similar to that for single DEMVP barrier (Figs. \ref{refrac} and \ref{demvpT}). At $3$T and near zero $V$, a forbidden region opens up at various values of the energy and is much larger than at $0.1$T. 
This can be seen in Fig.\ref{emvpbandky}. The gapped regions to the left and right of zero $V$ at $3$T are located in a pronounced asymmetric manner as compared to when $B=0.1$T, an effect also seen in Fig.\ref{demvpT}. 

This asymmetry in the band structure as a function of $V$ as well as the opening up of large forbidden zones for certain values of $V$ differentiates the transport through EMVP barriers from MVP barriers. Thus, it provides more flexibility to  tune  transport and constitutes one of the main findings of this work. 

\section{Conclusion}

We have analyzed in detail a new regime of coherent ballistic transport in monolayer graphene effected by the simultaneous application of highly inhomogenous magnetic fields alongwith electrostatic potentials. It has been found that the transport properties through a singular magnetic barrier can be much better controlled by the additional application of such electrostatic voltage. 
A detailed optical analogy for transport through single and multiple EMVP barriers arranged periodically was obtained highlighting optical analogues of phenomena such as TIR for positive and negative refraction and a Quantum Goos-H\"anchen shift. The most significant effect is that transport in this regime is highly anisotropic and strongly dependent on the sign of the voltage indicating possible device applications. 

All calculations have been done in the close vicinity of Dirac point with the assumption that the $K$ and $K'$ valleys are degenerate. The effect of disorder as well as interaction can be ignored assuming coherent ballistic transport.
Present techniques have indeed rendered such a regime experimentally accessible  and their scope is widening further \cite{Kim1}.  When the electron transport is exposed to a periodic arrangement of such EMVP barriers 
the band structure gets strongly modified. We have shown how this modification of band structure can be 
understood through an optical analogy. To explore the full band structure, one needs to calculate the impact of such periodic EMVP barriers within the tight binding approximation. However, tight binding calculations must take into account the finiteness of the barriers and it is important to note that the lattice spacing in graphene is generally much smaller than the width of the highly localized magnetic barriers.  A comparison of periodic structures of finite magnetic barriers \cite{Martino} shows that many properties related to transmission are very similar to what we have obtained in the delta function approximation. 

Since the above analysis is strongly affected by finite size effects \cite{Xu}, it will be interesting to include them by repeating our calculations for graphene nanoribbons. To summarize, our analysis of low energy transport near the Dirac point, provides many relevant results for graphene electronics \cite{Castro2, Geimstatus}.

\acknowledgments
We thank Prof. C. W. J Beenakker, S. Bhattacharjee and M. Barbier  
for helpful comments. SG also thanks Prof. R Moessner for his kind hospitality at MPIPKS Dresden, where this work was carried out partly. We acknowledge financial support by the IRD Unit, IIT Delhi and by DST, Ministry of Science and Technology.

\end{document}